\documentclass[sigconf,nonacm]{acmart}

\usepackage{booktabs} 
\usepackage{tikz}
\usepackage{graphicx}
\usepackage{bm}
\usepackage[export]{adjustbox}
\usepackage{color}
\usepackage{comment}

	\let\labelindent\relax
\usepackage{enumitem}
\usepackage[normalem]{ulem}
\usepackage{pbox}
\usepackage{hyperref}
\usepackage{multirow}
\usepackage{makecell}
\usepackage[caption=false]{subfig}
\usepackage{mathtools}

\usepackage{algpseudocode}
\usepackage{color}
\usepackage{array}

\usepackage{subfloat}
\usepackage{amsfonts}

\usepackage[vlined,ruled]{algorithm2e}

\graphicspath{results/plots}

\algblock{ParFor}{EndParFor}
\algnewcommand\algorithmicparfor{\textbf{parallel for}}
\algnewcommand\algorithmicpardo{\textbf{do}}
\algnewcommand\algorithmicendparfor{\textbf{end\ parallel for}}
\algrenewtext{ParFor}[1]{\algorithmicparfor\ #1\ \algorithmicpardo}
\algrenewtext{EndParFor}{\algorithmicendparfor}

\newcolumntype{L}[1]{>{\raggedright\let\newline\\\arraybackslash\hspace{0pt}}m{#1}}
\newcolumntype{C}[1]{>{\centering\let\newline\\\arraybackslash\hspace{0pt}}m{#1}}
\newcolumntype{R}[1]{>{\raggedleft\let\newline\\\arraybackslash\hspace{0pt}}m{#1}}
\usepackage{boxhandler}
\renewcommand{\baselinestretch}{0.98}

\newcommand{\HG}{HashGraph~}
\newcommand{\HGVOne}{HashGraph-V1}
\newcommand{\HGVTwo}{HashGraph-V2}
\begin{document}

\title{HashGraph - Scalable Hash Tables Using A Sparse Graph Data Structure}


\author{Oded Green}
\affiliation{%
	  \institution{NVIDIA} 
      \institution{Computational Science and Engineering, Georgia Institute of Technology}
}


\settopmatter{printacmref=false}
\setcopyright{none}
\renewcommand\footnotetextcopyrightpermission[1]{}
\pagestyle{plain}

\begin{abstract}

Hash tables are ubiquitous and used in a wide range of applications for efficient probing of large and unsorted data. If designed properly, hash-tables can enable efficients look ups in a constant number of operations or commonly referred to as $O(1)$ operations. As data sizes continue to grow and data becomes less structured (as is common for big-data applications), the need for efficient and scalable hash table also grows. 
In this paper we introduce HashGraph, a new scalable approach for building hash tables that uses concepts taken from sparse graph representations---hence the name HashGraph. We show two different variants of HashGraph, a simple algorithm that outlines the method to create the hash-table and an advanced method that creates the hash table in a more efficient manner (with an improved memory access pattern). 
HashGraph shows a new way to deal with hash-collisions that does not use ``open-addressing'' or ``chaining'', yet has all the benefits of both these approaches. HashGraph currently works for static inputs, though recent progress with dynamic graph data structures suggest that HashGraph might be extended to dynamic inputs as well.
We show that HashGraph can deal with a large number of hash-values per entry without loss of performance as most open-addressing and chaining approaches have. Further, we show that HashGraph is indifferent to the load-factor. Lastly, we show a new probing algorithm for the second phase of value lookups.
Given the above, HashGraph is extremely fast and outperforms several state of the art hash-table implementations. 
The implementation of HashGraph in this paper is for NVIDIA GPUs, though HashGraph is not architecture dependent. Using a NVIDIA GV100 GPU, HashGraph is anywhere from 2X-8X faster than cuDPP, WarpDrive, and cuDF. HashGraph is able to build a hash-table at a rate of 2.5 billion keys per second and can probe at nearly the same rate.

\end{abstract}

\maketitle

\section{Introduction}
\label{sec:intro}

Hash tables can be used for inner join \cite{blanas2011design}, group-by and aggregation \cite{karnagel2015optimizing}, query processing \cite{bethel2008bin}, looking up kmers in DNA sequencing \cite{pan2018optimizing}, intersecting voxel geometric objects \cite{alcantara2009real}, and counting triangles in graphs \cite{bisson2017high}. In some cases a dynamic hash-table is needed though for many of these applications a static hash-table is enough (where static refers to the fact that the input data is known a-priori before the construction of the hash-table).

Hashing is a major building block for numerous applications requiring lookups, such as associative arrays. One widely used case is for set intersection, which are also used in database join and group-by operations.
While many of these lookup operations can be implemented by first sorting the data, this can in fact lead to significant overheads. Thus, when the input data is sorted it is quite likely that the sorted intersection will outperform building a hash-table followed by probing. However, when the data is not sorted, which is very common, then hashing can prove to be more effective.
In fact, the question of which method, sorting vs. hashing, is preferable is one that has received significant attention \cite{kim2009sort}. To some degree, this question still remains open with new algorithms being developed on both fronts to improve the performance of both. For sorting, this includes new scalable algorithms which have better load-balancing, use more threads, and fully utilized new vector instruction sets. For hashing, new algorithms find better ways to reduce the number of conflicts, improve spatial locality while removing random accesses, and reducing storage overhead. 

In this paper, we present \emph{{\bf HashGraph}}. HashGraph is a scalable hashing algorithm for massively multi-threaded systems. We show two simple algorithms for creating a hash-table based on the widely used CSR (Compressed Sparse Row) data structure. CSR is used for representing sparse graphs and matrices---hence the term ``graph'' in HashGraph. Specifically, we show that it is possible to create a hash-table by visually thinking about the result of the hash as a graph. 

\subsection*{Contributions}
In this paper we introduce two novel algorithms for creating a static hash table. Both these algorithms are simple, straightforward, and give the same output.  
The first algorithm consists of five parallel \emph{for} loops and one parallel prefix operation. The second and more efficient algorithm consists of eight parallel \emph{for} loops and two parallel prefix operations. The straightforwardness of these algorithms, unlike many previous hashing algorithms, also means that they can be easily taught, adopted, and used by non expert programmers. Specifically, HashGraph is currently designed for applications requiring static data sets such as inner join, SpMV, and set intersections.

We show that HashGraph has all the benefits of chaining and open-hashing without their downsides. By using atomic instructions (namely an Atomic-Add operation), which is supported on most modern processors, we can build the hash-table by pre-processing the data. Specifically, in a first sweep across the data we are able to determine the exact number of elements that will be hashed to a specific index. With this information, we can then allocate the exact amount of memory for each index. This effectively removes our need to deal with collisions. Thus, we get a benefit similar to chaining but with much better locality (chaining requires random memory accesses). We get improved locality over open-hashing as the arrays have better locality and we are not required to iterate the table until an empty entry is found. 
Further, a unique feature of HashGraph is its insensitivity to load factor: we show that we do not need to increase the size of the hash-table to have double the number of input entries, as is typically required by open-hashing based approaches to ensure that insertion can be achieved in $O(1)$.

From a performance standpoint, HashGraph is able to hash inputs at a rate of 2.5 billion keys per second. This is almost twice as fast as several leading implementations for the simplest case where the input is the sequence of unique keys. When the keys are not unique and duplicates appear multiple times, then HashGraph can be as much as $40\times$ faster. HashGraph also closes the performance gap between hash based implementations and sorting based implementations. Using HashGraph we introduce a new method for hash table probing and set intersection that can work at nearly the same rate as our table building.

\section{Background}
\label{sec:related}

Hash tables are used across a wide range of application for efficiently looking up the existence of values in unsorted data. Given an input array of length $N$, building a hash-table typically takes $O(N)$ of operations. Given a good hashing function, ideally looking up a value takes $O(1)$ operations. In contrast sorting the input can take either $O(N)$ or $O(N \cdot log (N))$ operations (depending if a radix-sort or merge-sort is used, respectively). Looking up values, using a binary search, then requires an $O(log(N))$ operations. These theoretical bounds discuss the number of operations needed to create the data structure used for looking up values, but they do not cover the amount of locality in the memory accesses or the available parallelism. These tend to be dependent on a specific algorithm and implementation.

\paragraph*{\bf Hashing}

As a preliminarily we define a hash function as follows: 
\begin{equation}
	hash(a)\in \mathbb{N} \rightarrow b\in \left \{ 0,..,N/C_V -1 \right\}
\end{equation}, where $N$ is in the length of the input.

These functions take the original input and transforms them to numbers in the range of $\left \{0,..,N/C_V \right \}$. These values are the entries of the hash-table. In practice, most $hash(a)$ functions return values in $\mathbb{N}$, yet these values are divided by $N/C_V$ and the remainder is taken to ensure that range is reduced to a reasonable size---crucial for efficient memory utilization.
Murmur hash \cite{appleby2008murmurhash} is one such function that meets this criteria. 

$C_V$ is typically referred to as the {\bf load factor} of the hash-table. The load factor is referred to as $C$, we have replaced this with $C_V$ as $V$ is used to represent ``vertices'' used by our HashGraph. To reduce the storage overhead of the hash-table, $C_V$ can be set to values close to 1. 
However, this causes many challenging performance problems, including collision management and performance problems. As such $C_V$ is typically set in the range of $[0.5,0.8]$. 

The following lists several approaches for dealing with conflicts:

\emph{Separate Chaining} - each entry in the hash table maintains a linked-lists of values that have been hashed to it. For small $C_V~=1$, the linked lists tend to get longer as there are fewer hash values and the likelihood of a collision increases. This reduces performance due to an increased number of linked-list traversals and random memory access.


\emph{Open Addressing} - hash-values are stored directly within the table itself. For each hashed value, the table is scanned from the initial hash value until an empty spot within the table is found. While open-addressing has better caching than chaining, finding an empty spot can be costly. Open addressing is extremely costly when the number of duplicates is high or if a hash-value is used by numerous input values. Thus, if a value in the input exists $d$ times, finding an empty spot for each requires $O(d^2)$ operations. To reduce the likelihood of collisions for differnt input values, the load factor is typically set to $C=0.5$.

\emph{Coalesced Hashing} - uses a mix of separate chaining and open addressing. Rather than using a linked list for storing the elements, entries in the hash-table are used for storing elements and an additional pointer is used for storing the location of the next entry in a given chain. This typically gives improved locality over chaining and it can reduce the number of lookups for finding an unoccupied entry. This comes at the cost of additional storage (for pointer storage). In practice, our new HashGraph algorithm is closer to coalesced hashing though it does not store additional pointers for finding the next empty entry.

\emph{2-Choice Hashing} -  is an approach that uses two different hash functions with two tables. The hashed value is placed into the table with fewer entries. This can lead to better balancing \cite{richa2001power}. 

\emph{Cuckoo Hashing} \cite{pagh2004cuckoo} is similar to 2-Choice Hashing and can be extended to use more than one hash functions. However, unlike 2-Choice Hashing when a collision occurs across all the hash-functions one of the previous hash values is removed from the table and hashed using a different hash function. This is repeated until there are no more collisions. Thus, collisions can be quite costly.

\emph{Minwise Hashing} - is an approach for reducing the storage cost by using fewer bits \cite{li2012gpu}.

\paragraph*{\bf Parallel Hashing Algorithms}

One of the first parallel hash-tables for the GPUS was first introduced in \cite{bethel2008bin}. The table is split into multiple smaller tables that can be managed within smaller shared memory. Not long after this, a newer GPU algorithm was introduced in  \cite{alcantara2009real} based on Cuckoo hashing. 
The cuDPP \cite{harris2007cudpp} library is based on \cite{alcantara2009real}. Like most Cuckoo hashing approaches, the table building phase is greatly impacted by the appearance of duplicate values and results in a longer build time.

Ashkiani \emph{et al.} \cite{ashkiani2018dynamic} show a scalable approach of hashing using dynamic hash-tables for the GPU called SlabHash. Specifically, SlabHash is based on blocked linked-lists such that the payload of each node of the linked-list can hold multiple keys (or key-value pairs). 
SlabHash's performance is nearly equal to that of  cuDPP \cite{harris2007cudpp}.

WarpDrive \cite{junger2018warpdrive} is a new hashing algorithm for the GPU. WarpDrive scales to multiple devices due to the multi-split algorithm that they have developed. Currently WarpDrive can only work on building hash tables where the input is unique. An extensive review of multi-split approaches for the GPU can be found \cite{ashkiani2016gpu}.


Stadium Hashing \cite{khorasani2015stadium} is another hashing algorithm designed for GPU systems. Specifically, Stadium Hashing can work both in-core and out-of-core. This means that the hash-table can be larger than the physical memory of the GPU, which is prohibitive in some instances. Stadium hashing is based on Cuckoo hashing. Its performance is similar to  cuDPP \cite{harris2007cudpp}. Stadium Hashing is at it best when the hash table is larger than GPU's memory.

In \cite{maier2016concurrent} the ``Folklore'' hashing algorithm for the CPU is presented. Folklore is able to build the hash table at a rate of 300 millions keys per second on a 24 core (48 thread) processor and outperforms Intel's Threading Building Block Library. Folklore has the ability to support dynamically growing tables, though this  brings a significant performance penalty. 
A hardware optimized hash-join algorithm is given \cite{balkesen2013main}. Specifically, this algorithm targets hardware features such as SIMD instruction and is able to build and probe a hash-table at 200 million keys per second. This was later improved to 700 million keys per second at its peak rate in \cite{balkesen2013multi}, though on average the rate was close to 450 million keys per second. While faster than Folklore \cite{maier2016concurrent} , which reports the build rate and not the join rate as \cite{balkesen2013main,balkesen2013multi}, these rates are still significantly slower than current hash-table build rates on current GPU systems.

In \cite{blanas2011design,} a thorough review of hash-join algorithms and optimizations for multi-threaded CPU processors is given. Many CPU implementations \cite{blanas2011design,kim2009sort,balkesen2013main,balkesen2013multi} focus on reducing the number of cache misses. These tend to use some sort of data partitioning which is part of a pre-processing phase. Both \cite{kim2009sort} and \cite{blanas2011design} give each thread its own set of partitions which will later be merged with the results of the partitions . Thus, as the number of threads and partition grows, this can prove to be expensive. 
Unlike the GPU which requires tens of thousands of software threads to keep the device busy, most modern CPU systems only require fewer than a hundred threads to keep the processor fully utilized. This leads to very different partitioning approaches as well as optimizations. Many CPU optimizations for hash tables are not effective on the GPU.


Lastly, several algorithms have been designed for scalable distributed hash tables that can be used across thousands of CPUs cores and processors. This includes \cite{pan2018optimizing} and \cite{barthels2017distributed}. Both were tested on a CPU cluster with four thousand CPU cores. 
These distributed joins used different hashing functions to manage collisions.

\section{\HG}
\label{sec:hg}

In this section we introduce {\bf HashGraph}, a new and scalable approach for building hash tables that uses a  sparse graph-like data structure for static data sets, specifically compressed sparse row (CSR). In our CSR representations the vertices represent the hash values and the edges point to the original (pre-hashed) input. HashGraph is especially efficient for the hash table building phase, though it can also be used for the probing phase.

We show two variations of the building phase using HashGraph. The first version, called \HGVOne, (Sec. \ref{sec:hg-v1}) introduces a simple version of the HashGraph concept and is implemented with a handful of parallel for-loops. Using this version we highlight key features of the HashGraph. We then present a second  optimized variation, denoted \HGVTwo, that reduces the number of cache misses and increases performance. We show that by adding an additional preprocessing phases and several additional parallel for-loops we improve the performance over the first algorithm by roughly $4\times$.
HashGraph can utilize conventional probing mechanisms, yet, the uniqueness of the sparse graph data structure enables us to introduce a new probing method that benefits from an increased amount of locality. Specifically, this probing mechanism is unique to HashGraph.



\begin{figure*}[t]
\centering
\subfloat[Simple Hashing]{\includegraphics[width=0.80\columnwidth]{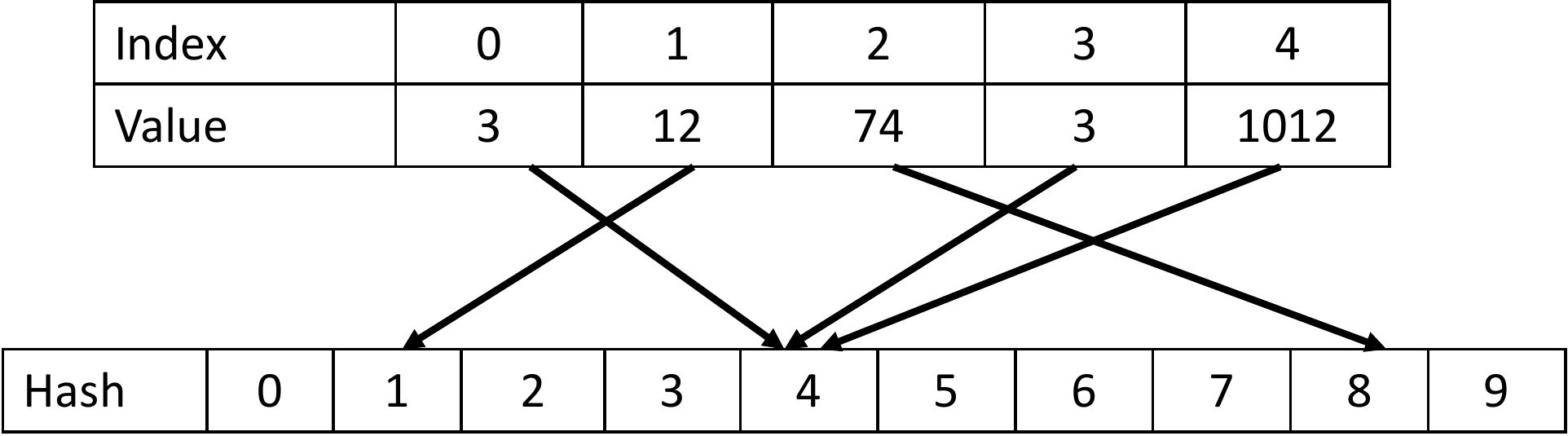}}
\hspace{1cm}
\subfloat[Hashing Using CSR]{\includegraphics[width=0.80\columnwidth]{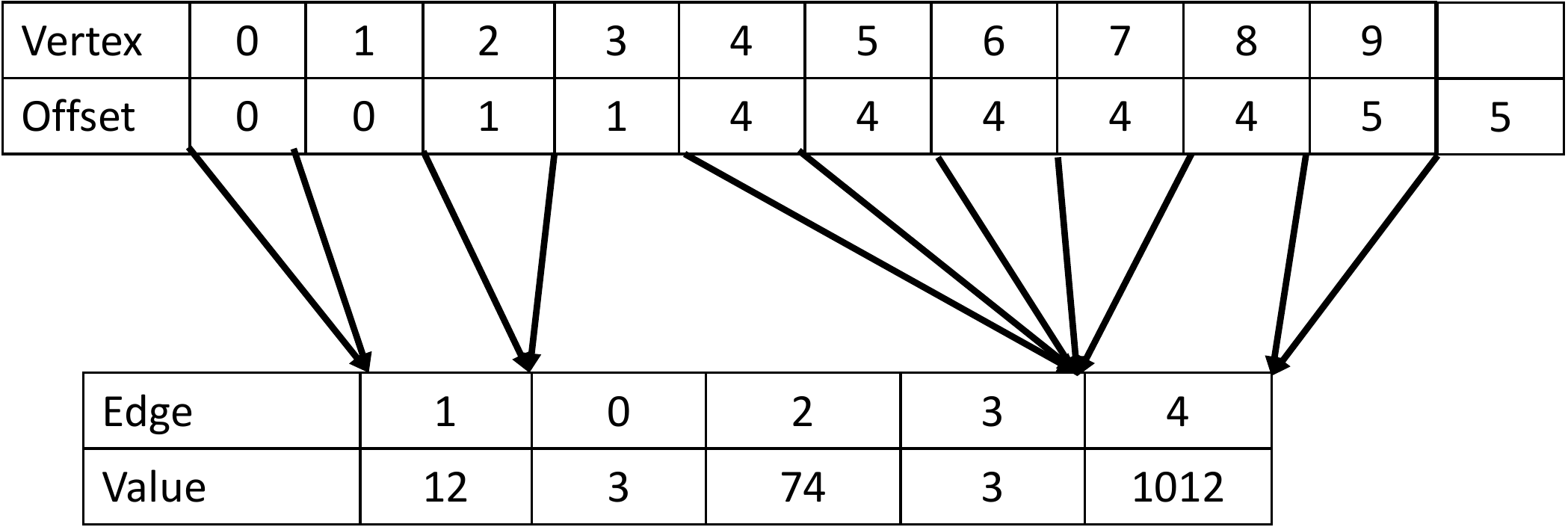}}

	\caption{(a) Input array with corresponding hash values. This is a bipartite graph with input values on one side and hash values on the other side. (b) Using a CSR-like data structure, a bipartite graph where the hash values point to the inputs.}
	\label{fig:simple-hashing}
\end{figure*}

\subsection{Hashing Using a Graph Data Structure}

As a first step, we show the relationship between hashing and graphs---this resembles a bipartite graph which can also be represented using CSR. The following subsections cover how to actually build the CSR-like data structure in an efficient manner whereas this section focuses on explaining why that representation is correct. Fig. \ref{fig:simple-hashing} (a) depicts an example of input array with 5 inputs that are hashed into a range of 10 values. In this example the number $3$ appears twice and is hashed to the same value as the value $10121$. This results in one of the hash values corresponding to three inputs. Note that the result of the hash is essentially a bipartite graph. Going from the input to the hash is trivial (simply use the hash function). The reverse direction is achieve by having the hash-values point back to the inputs that were hashed into them.

For most hash tables, where the input is of size $N$, the size of the hash range is $\frac{N}{C_V}$. Typically $C_V=0.5$ for open-addressing based methods (to reduce collisions and to ensure $O(1)$ access time). For Hash Graph, we will show in Section \ref{sec:performance} that $C_V=1$ gives better performance than $C_V=0.5$ despite common intuition that a lower load-factor makes it easier to hash and place keys into the hash.

\paragraph*{Compressed Sparse Row}
CSR is a widely used data structure to represent sparse datasets for both matrix and graph based problem. CSR stores only non-zero values of the input. For each row, the non-zero values are ``compressed'' such that the value and the column of the value are stored. In graph terminology, each of these rows is the adjacency array of a given vertex also referred to as the edges of the vertex. 
CSR uses a small number of arrays 1) for storing the vertex sizes (array length is $|V|+1$) and 2) an array for the edges (an array of size of $|E|$ elements). The vertex adjacency array sizes are in fact stored via an offset array which points to the starting point of a vertex's adjacency array. The end of the adjacency array is the offset to the next vertex. Thus, the size of a vertex's adjacency array is the difference in the position of two consecutive vertices: $v$ and $v+1$. The edges are stored as either a single array of tuples $(column, value)$ or via two arrays (one for columns and one for values). 
Fig. \ref{fig:simple-hashing} (b) depicts how to represent the reverse relationship from hash-values to the inputs using CSR for the example in Fig. \ref{fig:simple-hashing} (b). Note, some hash entries might not have inputs that are hashed into them, as such their adjacency array length is zero. 


\begin{algorithm}[t]
\DontPrintSemicolon

\scriptsize

\caption{HashGraph-V1 - simple algorithm for building hash table. Note $V=N/C_V$.}
\label{alg:hg1}

\ForPar{$i=0:1:N-1$}
 {
 // These are the vertices of the graph \;
 $  H_A[i] = hash(A[i]) \: mod \: V$ \;
 }

\ForPar{$i=0:1:V-1$}
 {
 $	CounterArray[i] \leftarrow 0 $ \;
 }

\ForPar{$i=0:1:N-1$}
 {
 $	AtomicAdd (CounterArray[H_A[i]], 1) $ \;
 }

// The offset array is not used in the computation. Rather it is stored for the probing phase s.t. the lengths of each hash table entry is known. \;

$ OffsetArray \leftarrow PrefixSum(CounterArray)$ 

\ForPar{$i=0:1:V-1$}
 {
 $	CounterArray[i] \leftarrow 0 $ \;
 }

\ForPar{i=0:1:N-1}
 {
 $	pos \leftarrow AtomicAdd (CounterArray[H_A [i]], 1) $ \;
 $E[pos] \leftarrow (A[i],i)$ \;
 }


\end{algorithm}

    

\subsection{HashGraph Version 1.0}
\label{sec:hg-v1}


In this subsection we introduce a simple approach for creating the HashGraph data structure. Pseudo code for this can be found in Alg. \ref{alg:hg1}. Note, that algorithm for building HashGraph  is fairly straightforward and consists of five simple parallel for-loops and one parallel prefix sum operation. In contrast many parallel hash-table implementations consist of either one large monolithic code section or an advanced parallel sections that includes numerous barriers and synchronizations. In most cases, it is the fact that these algorithms are so complex that they are designed for a specific processing system. For example, cuDF's \cite{cudf} hash table uses a single GPU kernel to build the a hash table. In contrast, HashGraph uses a large number of simple kernels. 
The following explains the process for building a HashGraph.


\noindent {\bf 1)} \  Each value in the input is hashed into the range of the hash-table. This operation is highly parallel and uses a sequential memory access pattern making it cache-friendly.

\noindent {\bf 2)} \ For each vertex in the graph, aka the set of possible hash values, the number of instances of that entry is set to zero. This phase is also cache friendly due to its sequential memory access.

\noindent {\bf 3)} \ For each of the hash values, computed 1) , we proceed to count the number of instances that each hash value appears in the original input. When this for-loop is completed, we know how many elements will belong to each vertex in the CSR data structure. 
This parallel for-loop loop is computationally more expensive than the first for-loop even though both loops iterate over an equal number of elements. Specifically, this for-loop uses atomic instructions for counting the number of instances and writes to random places in memory.  

\noindent {\bf 4)} \ A prefix sum array is computed. Parallel prefix sum operations are well studied \cite{blelloch1990pre,harris2007parallel,sengupta2007scan}.
The result of the prefix sum operation is the offset array found in CSR. Through this offset array we can now partition the edge array based on the number of values hashed to each entry. 

\noindent {\bf 5)} \ For each entry in the table, the number of instances of that array is reset to zero. This array is used in the final phase for maintaining the relative position of hashed elements with in the list that the have been assigned to. This phase is cache friendly due to its sequential memory access. Note, the $OffSetArray$ is stored and used for the probing phase.

\noindent {\bf 6)} \ In the last and final phase, the hashed values are actually placed in the HashGraph. Specifically, each hash-value is placed in the element list that it belongs to by getting the position from the offset array and the relative position within that array using atomic instructions. 

\paragraph*{Collision Management}
The performance of traditional hashing tables is highly dependent on the load factor and the hash range.
As traditional hash table becomes denser, the overhead of finding an empty spot in an open-addressing based hash table becomes more expensive. This is problematic for chaining based hash tables and will results in longer lists (``chains'') and will make probing significantly slower.

\paragraph{HashGraph Collision Management} Recall, the load factor for HashGraph is different than it is for open-addressing based methods. In HashGraph the load factor states how many ``vertices'' will be in the graph or what is the range of the hash values in the hash table and not the size of the table. In HashGraph the table size is exactly the size of the input size. 
Unlike open-addressing which requires searching for an empty spot in the hash table to find the end of a ``value-chain'', in HashGraph the entries for a specific hash value are in consecutive memory. Further, the length of that chain is known and as such only the relevant values are accessed. Lastly, as the HashGraph building phase includes a preprocessing for counting the number of entries hashed to a specific value, the performance of HashGraph does not degrade if an entry exists numerous times. This will be discussed in more detail in Sec. \ref{sec:perf-hg} (Fig. \ref{fig:dup-keys}).


\paragraph*{Work and Time Complexity For Building \HGVOne}
The following analyzes the work complexity for Alg. \ref{alg:hg1}. First, note all for-loops iterate over $N$ or $C_V$ elements. 
Only the first for-loop requires special attention as the work complexity is dependent on the cost of doing the hashing, namely $O(N \cdot Hash)$. For the remaining for-loops 
work complexity is either $O(N)$ or $O(\frac{N}{C_V})$. 
In \cite{harris2007parallel,sengupta2007scan} it was shown that a parallel prefix sum operation can be work efficient (linear amount of work with respect to the array size), $O(\frac{N}{C_V})$. 
This leads to a total time complexity of $O(N+\frac{N}{C_V}+N\cdot hash)$. 
When $Hash~O(1)$ (which is the case for many hash functions), the overall work complexity can be simplified to $O(N+\frac{N}{C_V})$.

Given $P$ cores and the above work complexity, we can evaluate the time complexity for HashGraph as follows. All five for-loops are embarrassingly parallel and are easily partitioned across $P$ cores, as their time complexities is either $O(N/P)$ or $O(\frac{N}{C_V\cdot P})$. 
The time complexity of the prefix operation is $O(N/P+log(P))$ \cite{harris2007parallel,sengupta2007scan}. As such the time complexity can be written as $O(N/P+\frac{N}{C_V\cdot P} +log(P))$.  In practice $log(P)$ will be significantly smaller than the other components and will be negligible for most inputs.

\paragraph*{Storage Complexity}

In the following storage complexity analysis we avoid using the standard $O()$ notation as the constants matter. Instead, we offer a complexity analysis that states the array sizes, in number of elements. For the sake of breivity, we do not consider the actual bit and byte representation. 
The size of the table is exactly $N$ elements. The table itself can store either only the unhashed value or the unhashed value and the index of that value in the input array (this is needed for join operations). Further, we use two additional arrays of size $\frac{N}{C_V}$ for the counter and the prefix sum array. In practice, HashGraph works well with $C_V=2$.

In contrast, other open-addressing hash tables typically require a large table with $\frac{N}{C_V}$ elements and $C_V=0.5$ for performance reason. Due to the difference values of $C_V$ used by the hash tables, HashGraph requires a similar number of elements.

\begin{algorithm}[t]
\DontPrintSemicolon

\scriptsize

\caption{HashGraph-V2 - cache efficient algorithm for building hash. Note $V=N/C_V$.}
\label{alg:hg2}


// Phase 1: Bin Counting

$BinSize \leftarrow \frac{V}{Bins} $;

\ForPar{$i=0:1:Bins-1$}
 {
 $  BinCounterArray[i] \leftarrow 0; $
 }

\ForPar{$i=0:1:N-1$}
 {
   $bin \leftarrow \frac{hash(A[i]) \, mod  \, V}{BinSize}; $  \;
 $  AtomicAdd (BinCounterArray[bin], 1); $ 
 }

// Phase 2: Bin Placement and Data Reorganization

$ BinOffsetArray \leftarrow PrefixSum(BinCounterArray);$

\ForPar{$i=0:1:Bins-1$}
 {
 $  BinCounterArray[i] \leftarrow 0; $ 
 }

\ForPar{$i=0:1:N-1$}
 {
    $bin \leftarrow \frac{hash(A[i]) \, mod  \, V}{BinSize}; $  \;
    $pos \leftarrow atomicAdd(BinCounterArray [bin],1)+ BinOffsetArray[bin]; $ \;
    $ReorgA[pos] \leftarrow \{A[i],i\}; $\;
 }

// Phase 3: HashGraph Creation

\ForPar{$i=0:1:V-1$}
 {
 $  CounterArray[i] \leftarrow 0; $ 
 }

\ForPar{$i=0:1:N-1$}
 {
    $pos \leftarrow hash (ReorgA[i].val) \, mod\, V  ;$ \;

    $atomicAdd (CounterArray[pos], 1); $ \;
 }

// The offset array is not used in the computation. Rather it is stored for the probing phase s.t. the lengths of each hash table entry is known. \;

$ OffsetArray \leftarrow PrefixSum(CounterArray);$ 

\ForPar{$i=0:1:V-1$}
 {
 $  CounterArray[i] \leftarrow 0; $ \;
 }

\ForPar{i=0:1:N-1}
 {
 $  pos \leftarrow AtomicAdd (CounterArray[H_A [i]], 1); $ \;
 $E[pos] \leftarrow (A[i],i)$; \;
 }




\end{algorithm}

    

\subsection{HashGraph Version 2.0}
\label{sec:hg-v2}
We introduce a modified version of HashGraph that is more memory subsystem friendly: cache efficient and with improved memory locality. HashGraph-V2 (Alg. \ref{alg:hg2}) increases the number of parallel for-loops from five loops to eight loops and uses two parallel prefix sum operations instead of just one. Yet, all these for-loops are friendly to the memory subsystem and have good spatial and temporal memory accesses. HashGraph-V2 consists of three phases, two of them reorganize the input in a cache friendly manner. 

\paragraph{Intuition for two phase of data movement} - in a preliminary phase, we count the number of hash values that fall in a specific range. This is done by binning hash values together. 
The number of bins is dictated by the bin-size and the cache size. Ideally, the number of elements in each bin (``bin size'') should also fit into the cache. Thus we want to avoid small bins (equivalent to HashGraph-V1) or a large bins (that are too large to be cached).
In practice, for hash tables with hundreds of millions of entries\footnote{We tested up to 512M entries} we found that a good number of bins is roughly 32k. For these sizes, the bins on average have a few hundreds of thousands of elements and as such each bin can fit in the last level cache (referred to as the LLC). This allows us to bring one bin at a time in the LLC and then reorganize the bin efficiently within the cache.


The three phases found in Alg. \ref{alg:hg2} are as follows:

	\noindent {\bf Phase 1}: Bin Counting - given an array of $Bins$ where each each bin is of size $BinSize$, we count the number of elements hashed into each range. Specifically, the bin ranges are from $(0...BinSize-1)$, $(BinSize... 2\cdot BinSize-1)$, $...$, and so forth. Also, the array $Bins$ is cache-able making this operation efficient. Once the hash values have been counted, a prefix is computed across the bins. 
 	The parallel for-loops in this phase have good spatial locality.

	\noindent {\bf Phase 2}: Bin Placement and Data Reorganization - the original input $A$ is re-ordered into a new array $ReorgA$. The new array is a tuple of two values: the value of $A[i]$ and its index $i$ (needed for the hash-table itself)---this doubles the memory footprint; however, it improves the spatial and temporal locality of the memory accesses. 
	Note, as the number of $Bins$ is relatively smaller, the cache line that each of these bins points to is also cache-able---this means that writing the values of $ReorgA$ is cache efficient.
	
	\noindent {\bf Phase 3}: HashGraph Creation - the final phase takes the reorganized data and creates the HashGraph in a second sweep over the data. The data is moved into its final position by counting the number elements that will be hashed into each element. Similar to the previous phase, this phase also has good spatial and temporal locality.

\paragraph{Larger Inputs}
For the sake of simplicity in HashGraph-V2 and its pseudo code (Alg. \ref{alg:hg2}), we focused on only one iteration of relabeling. However, for data sets with billions and trillions of records in the input it is very likely that either the number of bins needed for the efficient relabeling will be so high that it cannot fit into the LLC. As it might be necessary to use multiple iterations of relabeling, thought its quite likely for the trillion of inputs no more than two or three phases of re-ordering will be needed or perhaps a cache oblivious algorithm. Such approaches have already been investigated for large shared-memory systems.
This paper does attempt to solve this problem, yet we do not ignore the fact that this could be a problem for larger inputs. We think that a multiple tier HashGraph could be useful.


\paragraph*{Work and Time Complexity For Building \HGVTwo}

For the sake of brevity, we reuse the analysis used for the first algorithm. First note that all the for-loops go over: $Bins$, $N$, or $\frac{N}{C_V}$ elements. The complexities of these loops are identical to those found in the HashGraph-V1 algorithm. While the number of for-loops has increased, we can still write the work complexity as: $O(Bins+N+\frac{N}{C_V})$. As $Bins<<N$ we can rewrite the work complexity as  $O(N+\frac{N}{C_V})$ similar to the first algorithm.
The time complexity is $O(N/P+\frac{N}{C_V\cdot P})$, similar to the first algorithm. 

\paragraph*{Storage Complexity}

HashGraph-V2 has a larger memory footprint that HashGraph-V1. Specifically, the relabeled array doubles it foot print.

\begin{algorithm}[t]
\DontPrintSemicolon

\scriptsize

\caption{HashGraph-Probe-New - New hash table probing algorithm that uses two HashGraph for the probing. $IntersectAdjacencyArray$ receives two lists and their lists. }
\label{alg:probing}

// Create two \HG \; 
 $  HG_A \leftarrow HashGraph\_V2(A, N_A,V); $\;
 $  HG_B \leftarrow HashGraph\_V2(B, N_B,V); $\;

// Intersection
\ForPar{$i=0:1:V-1$}
 {
  $IntersectAdjacencyArray(HG_A[i],HG_B[i])$ \; 
 }

\end{algorithm}

\subsection{Probing Using HashGraph}
\label{sec:hg-probe}




The probing process in a hash-table is the phase in which the values from a second input are looked up within the hash table created from the first input. In the case of set intersection, all the values of the second set are looked up inside the hash-table to find common elements. Give the fact that HashGraph shares structural similarities with open-addressing, this enables using the  probing mechanism of open-addressing for HashGraph. We refer to this approach as $HashGraph-Probe-Standard$.
An interesting artifact of using this probing mechanism with HashGraph is that it will have better performance with HashGraph than it will with open-addressing. Specifically, 
HashGraph will iterate over fewer entries than open-addressing (as different hash-values can create a longer chain of entries within the table due to local collisions). 

The uniqueness of HashGraph allows designing a new and cache efficient probing mechanism. We refer to this approach as $HashGraph-Probe-New$. This approach can be especially efficient for join operations in databases. Pseudo code for our new probing mechanism can be found in Alg. \ref{alg:probing}---note the simplicity of this probing algorithm. 
Specifically, given two input sets we create two HashGraphs---one for each input. We can the proceed to intersect the adjacency arrays for the corresponding lists in each of the table. The benefit of doing these list intersections is cache locality. For simplicity, assume that the first hash-table, $A$, has a set of elements $A_x={A_{x,1}, A_{x,2}, A_{x,3},...}$ that have been hashed to the value $x$. Now assume that the input $B$ has the following $B_x={B_{x,1}, B_{x,2}, B_{x,3},...}$ set of values of hashed to $x$. By using $HashGraph-Probe-Standard$, for each element in $B_x$, the elements of $A_x$ need to be fetched in the memory. In contrast, by using $HashGraph-Probe-New$ we can fetch in $A_x$ only once into the cache and reuse that data. In Sec. \ref{sec:performance} we show that the cache reuse is substantial. We believe that the new probing function will be useful some applications such as SpGEMM where the cost of building the second hash table gets amortized by re-use. Specifically, consider the need to intersect multiple rows and columns in a sparse setting. The HashGraph for each of these rows or columns needs to be built once but can be used in multiple intersections.

With HashGraph-V2, we showed how to create the hash table in a cache efficient manner. Thus, the new probing approach is also cache efficient. The benefit of our new probing approach is that if we can further improve the building phase, then probing itself can also become faster and less dependent on random memory accesses. 

The trade off of $HashGraph-Probe-New$ is that we need to build an additional HashTable to get is performance. Thus, the decision on using $HashGraph-Probe-Standard$ vs. $HashGraph-Probe-New$ could very much depend on the data and the application and available storage. In Sec. \ref{sec:performance}, we show several instances that $HashGraph-Probe-New$ is faster in the context of join operations.

\section{Experimental Setup}
\label{sec:experiments-setup}

\paragraph{System and Configuration}

Our implementation of HashGraph is in CUDA and targets NVIDIA'sGPU. 
The GPU used in this experiment is the NVIDIA Quadro GV100.
The GV100 is a Volta (micro-architecture) based GPU with 80 SMs (streaming multi-processors) and 64 SPs (streaming processors) per SM, for a total of 5120 SPs. The SPs are lightweight hardware threads that are sometimes referred to as CUDA threads. In practice, roughly 40K software threads are required for fully utilizing the GPU.
The GV100 has a total of 32GB of HBM2 memory and 6MB of shared-cache between the SMs. Each SM also has a configurable shared memory of $96KB$. The GV100 has 640 tensor cores, though these are not used in our experiments. The GV100 has two variants, PCI-E and SXM2, form factors. 
Our GV100 is the PCI-E variant that has a peak power consumption of 250 watts.
The GV100 is connected to an Intel $i7-7800X$ (6 cores with 12 threads) running at 3.5 GHz with a 8.5MB last-level cache.

Lastly, to benchmarking the hashing functionality on a CPU system we use a dual-processor Intel Xeon Platimun 8168 system. This system has 48-cores with 96 threads with a frequency of 2.7GHz, and 1.5 TB of DRAM.

\paragraph{Frameworks}

As our implementation targets the GPU, we primarily compare against other leading GPU implementation though we also compare to one of the fastest CPU based hash tables used for inner joins. To compare fairly against all the GPU implementations, we benchmark the table creation and table probing processes separately. We do this as the GPU implementations clearly distinguish between these phases in the implementations. Further, this allows better performance analysis to see if one of the phases is a performance bottleneck.
For the CPU implementation found in \cite{balkesen2013main,balkesen2013multi}, which was designed for inner joins, we were not able to separate the building and probing phases, as such we compare the total time.

\emph{\bf cuDPP \cite{harris2007cudpp}} is one of the first hast table implementations for the GPU and is also one of the best performing. cuDPP uses a Cuckoo based hashing and uses open addressing. The hash table built by cuDPP is static. It does not support insertions or deletions.

\emph{\bf WarpDrive \cite{junger2018warpdrive}} is a GPU specific hash table that gets good performance by utilizing several hardware features such as warp synchronization operations and warp groups. Note, WarpDrive currently only works with hash-tables were all in the values in the input are unique---this is a big restriction for real data sets. The hash table built by WarpDrive is static. It supports insertions but does not support deletions (as this requires adding a special marker for deleted elements.

\emph{\bf cuDF \cite{cudf}} is a framework for GPU dataframes. cuDF supports a wide range of ETL (Extract, Transform, Load) functionality on the GPU. cuDF uses a linear based open addressing approach for its hash table. The hash table built by cuDPP is static. It does not support insertions or deletions.

\emph{\bf MC-Hash \cite{balkesen2013main,balkesen2013multi}} MC-Hash (multi-core hashing) is one of the fastest hash-tables on CPU systems. Specifically, MC-Hash targets join operations commonly used in data-bases, though it can also be utilization for other use-cases.

The GPU implementations use open address based tables. As such their performance is highly dependent on the load factor. By default we set the load factor to $50\%$ and allocate twice as many entries in the hash table then those found in the input.
For the probing phase, we focus on probing the hash table with a secondary input array (similar to the process found in joins or list intersections). Note, several of these frameworks support insert operations into the hash-table before it fills up. They do not support deletions. In contrast, HashGraph can easily support deletions by storing the number of elements per hash. See Sec. \ref{sec:dynamic-hg} for a discussion for insertion operations for HashGraph.

Chaining based hash table on the GPU are challenging due to memory management and performance. The recent SlabHash \cite{ashkiani2018dynamic} is the closest to a chaining based hash-table on the GPU; however, we do not compare against SlabHash \cite{ashkiani2018dynamic} as it is out performed by cuDPP \cite{harris2007cudpp}.

\paragraph{Input sizes \& Key Distribution}

In many data base application where hash-table are used, typically one of the tables is much smaller than the other. For data sciences, it is customary to have inputs with similar dimensions. For the sake of simplicity, we primarily focus on the later case. 
We report performance in throughput (keys per second) rather than absolute times as is common with hash table operations. As the table creation and probing processes are separated, times can be inferred for both these operations based on the input size divided by the throughput.


By default, both input arrays will have 32M keys unless mentioned otherwise and the input across the frameworks is identical ($[1,2,3,4,..., 2^{25}]$) to ensure fair evaluation across the implementations. We report for 32M keys as we found that several of the frameworks were unable to scale to large data sets. We chose the simple sequence of number as it ensures that all  frameworks get the same input, especially as WarpDrive \cite{junger2018warpdrive} is unable to deal with duplicate values.

For HashGraph and cuDF we test the performance across a wider range of input sizes as well as random inputs (controlling the average number of appearances of each value in the input).
For HashGraph, we were able to build a hash table on a single GPU with $2^{29}=512M$ entries.
Beyond that size, we ran out of memory. From a performance perspective, HashGraph's throughput is  roughly within a $5\%$ range of its performance with $32M$ entries as the input is increased to $512M$ entries.

\section{Empirical Performance}
\label{sec:performance}
\begin{figure*}[t!]

\centering
\subfloat[Keys Per Second]{\includegraphics[width=0.83\columnwidth]{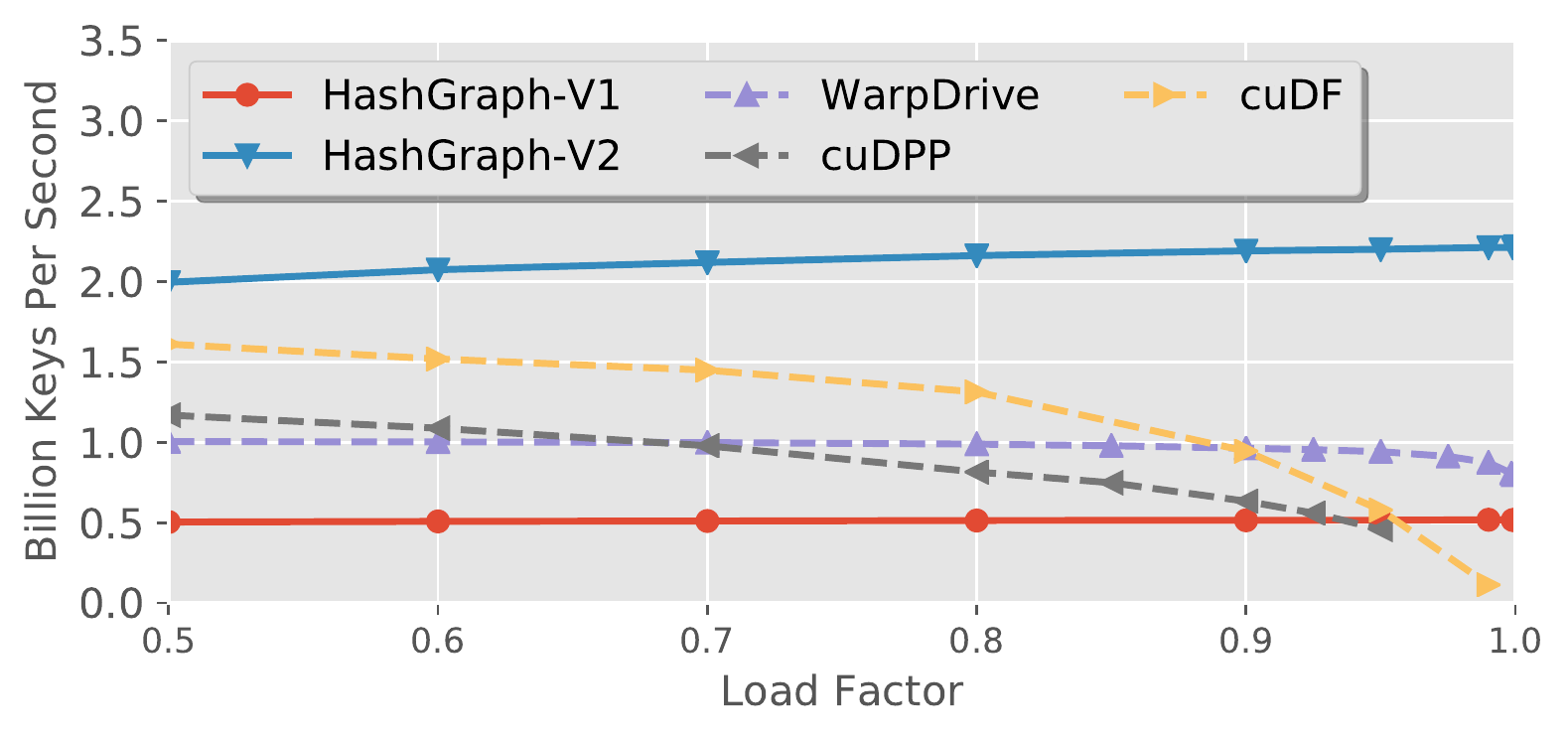}}
\hspace{1cm}
\subfloat[Normalized Speedup]{\includegraphics[width=0.8\columnwidth]{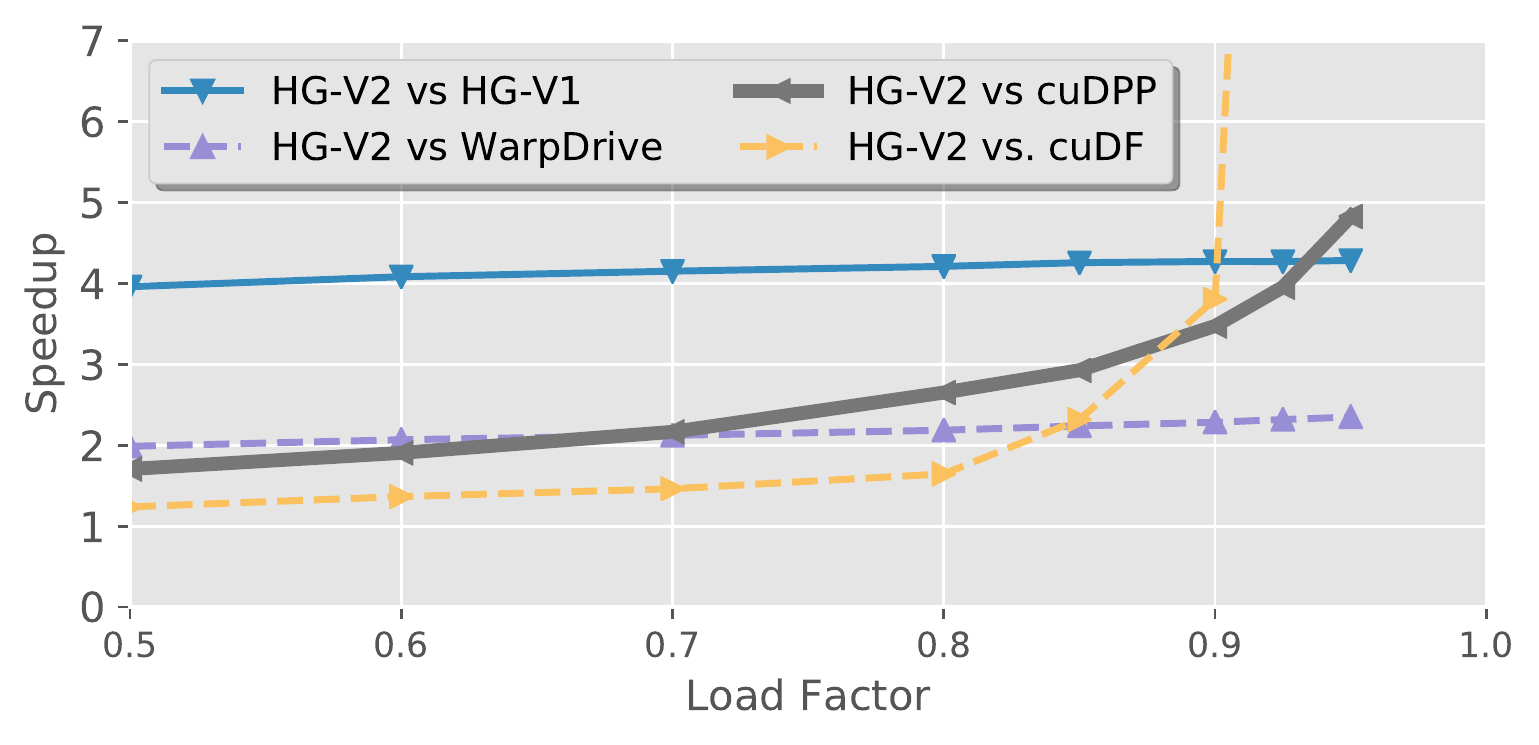}}

\caption{Performance of building a hash-table. X-axis represents the load factor of hash table and y-axis represents the (a) billion of keys per second (higher is better) and (b) the normalized speedup in comparison with cuDPP (higher than is better performance). Input for all table is identical and is the sequence $[1,2,3,...,2^{25}]$.
For HashGraph, increasing the load factor actually reduces the number of vertices improves the performance of the hash-table creation. This is in contrast to the typical behavior of hash-table creation.
}

\label{fig:hash-graph}
\end{figure*}

\begin{figure}[t!]
\centering
\includegraphics[width=0.8\columnwidth]{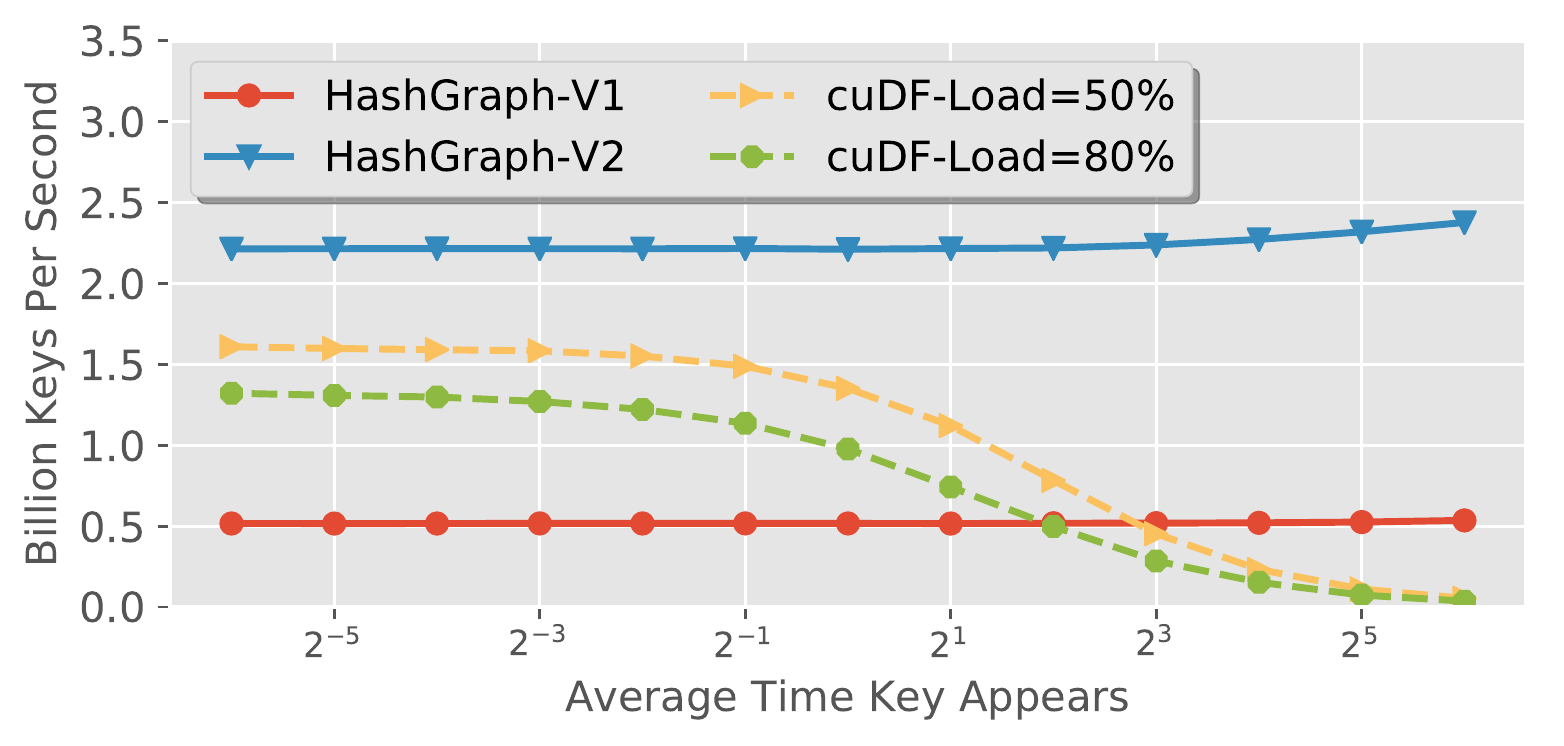}
\caption{Billion of keys per second (higher is better) for building a hash table HashGraph and cuDF with two load-factors as a function of the average number appearance per key in the input. 
Keys are generated using a random number generator and a predetermined key range. Everything to the right of $x=0$, the key range is smaller than the number of inputs $N/R$ meaning that the average number of appearances grows with range $R$. Left of the $x=0$, the key range is $N \cdot R$, meaning that the average time a key appears is smaller. 
The performance of HashGraph is almost indifferent to the number of appearances whereas for cuDF the performance deteriorates with the increase.
When the average number appearances is $8\times$, the performance of HashGraph is anywhere from $4\times$ to$40\times$ faster.}

\label{fig:dup-keys}
\end{figure}

\begin{figure*}[t]

\centering

\vspace{-0.1cm}
\subfloat[Input size - $2^{21}$]{\includegraphics[width=0.60\columnwidth]{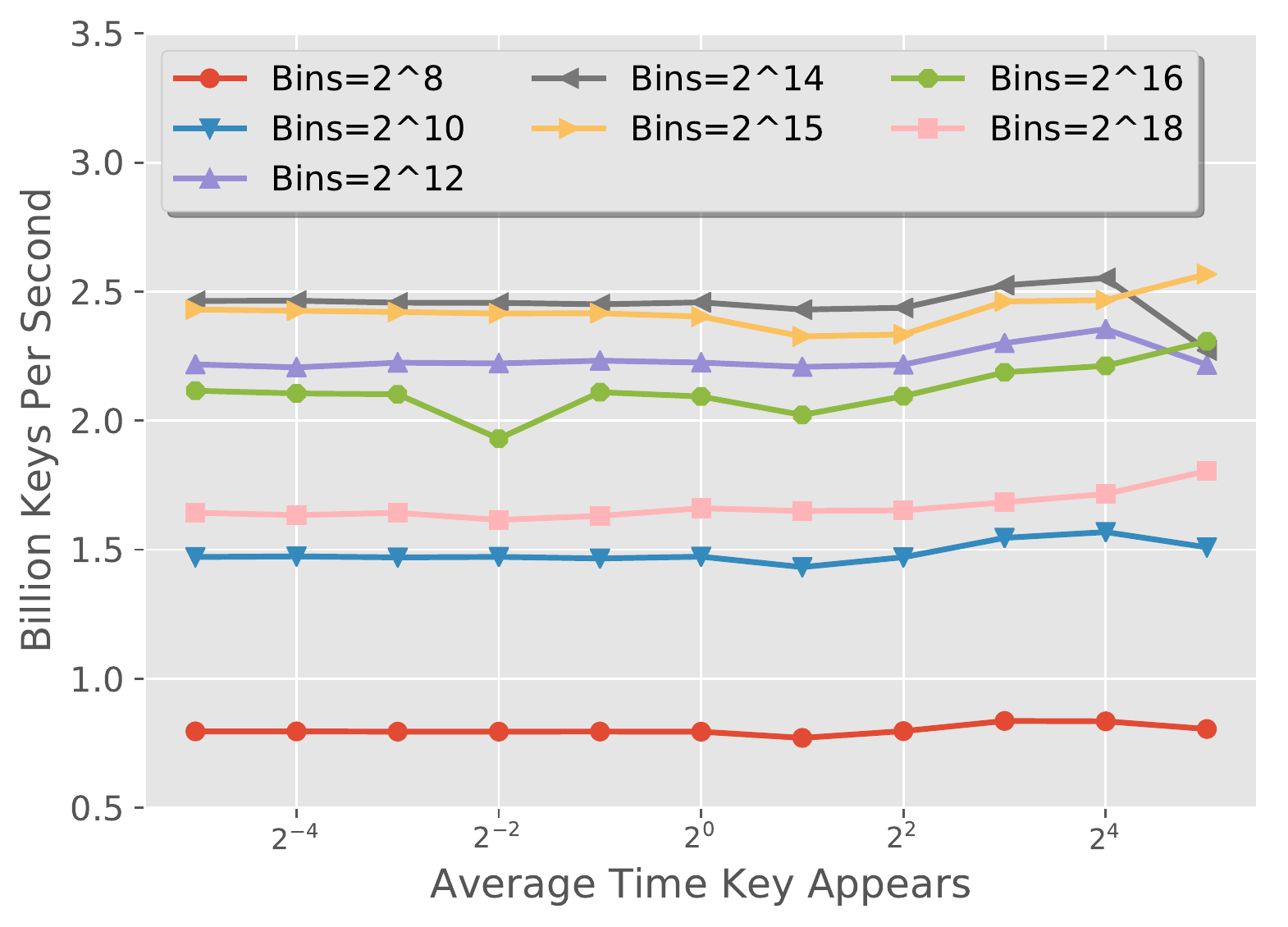}}
\hspace{0.5cm}
\subfloat[Input size - $2^{23}$]{\includegraphics[width=0.60\columnwidth]{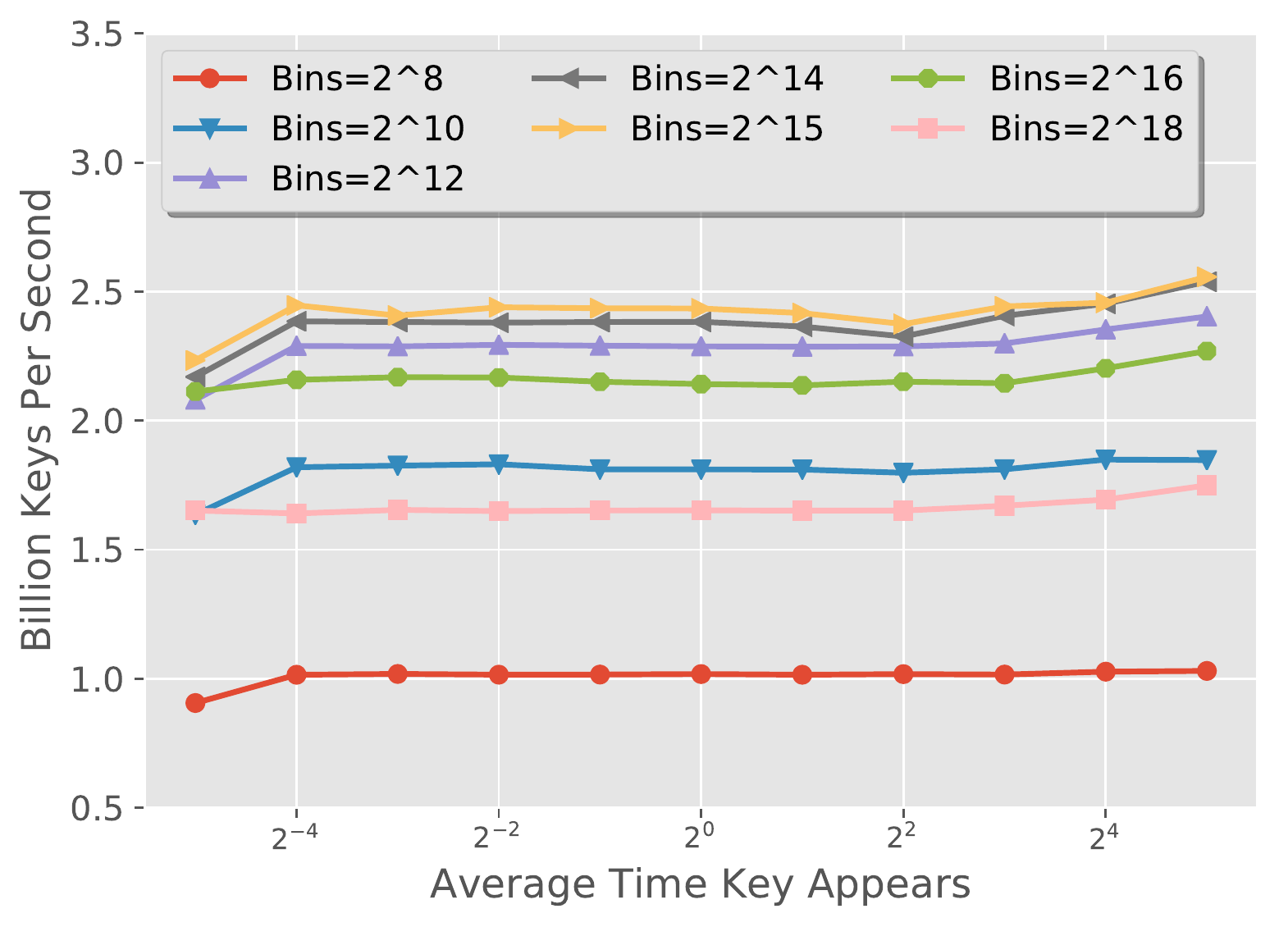}}
\hspace{0.5cm}
\subfloat[Input size - $2^{25}$]{\includegraphics[width=0.60\columnwidth]{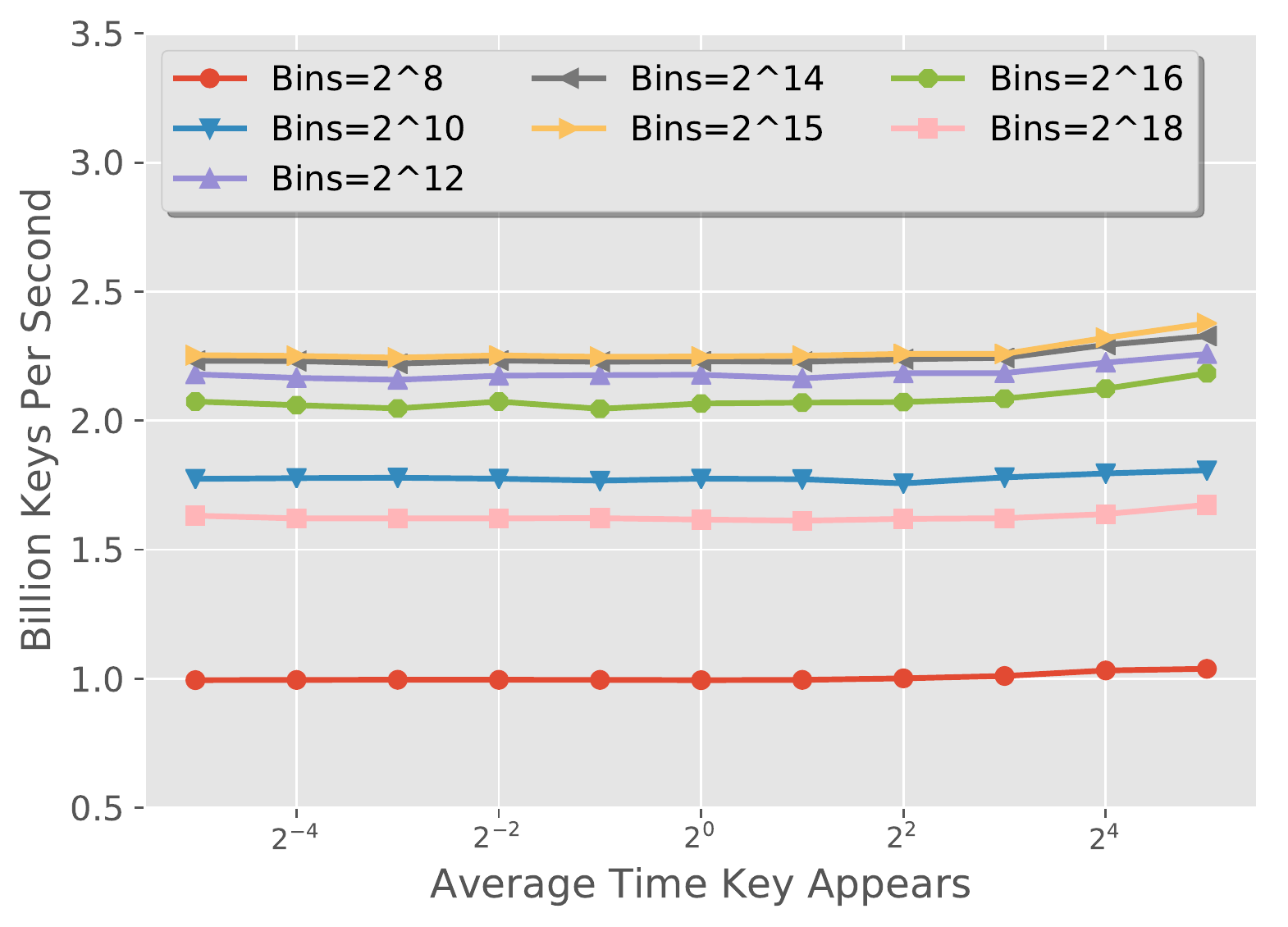}}

\caption{Billions of keys per second of our HashGraph 2.0 as a function of the number of times each key appears in the input. The size of the input is (a) $2^{21}$, (b) $2^{23}$, and (c) $2^{25}$ elements, respectively. The different curves are for a different number of bins. Best performance is achieved for $2^{14}$ and $2^{15}$  bins due to caching. At peak, HashGraph is able to build the hash table at a rate of 2.5 billions keys per second.
}
\label{fig:size-vs-bins}

\end{figure*}

\begin{figure*}[t]

\centering
\vspace{-0.2cm}
\subfloat[Bins - $2^{14}$]{\includegraphics[width=0.60\columnwidth]{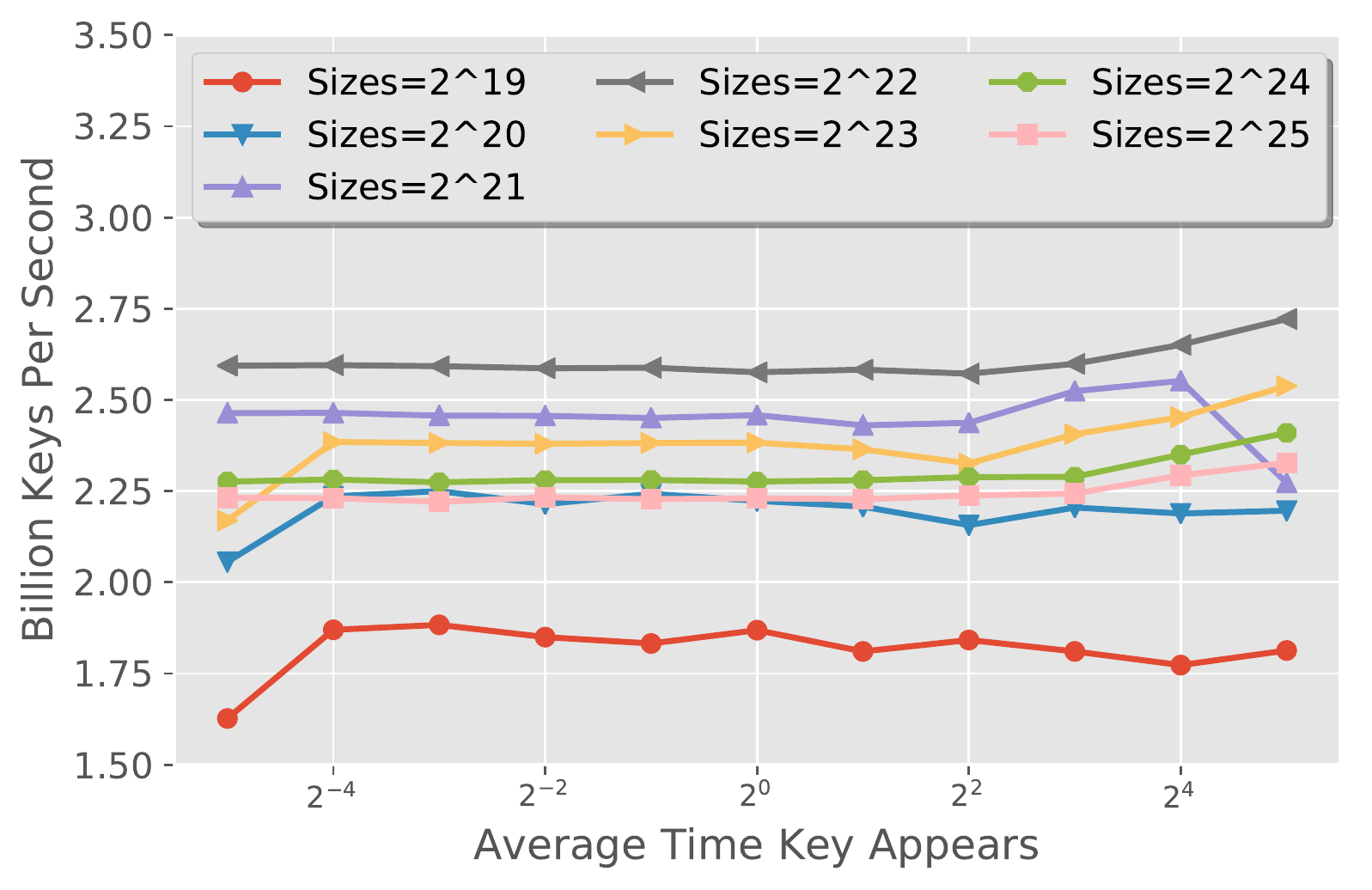}}
\hspace{0.25cm}
\subfloat[Bins - $2^{15}$]{\includegraphics[width=0.60\columnwidth]{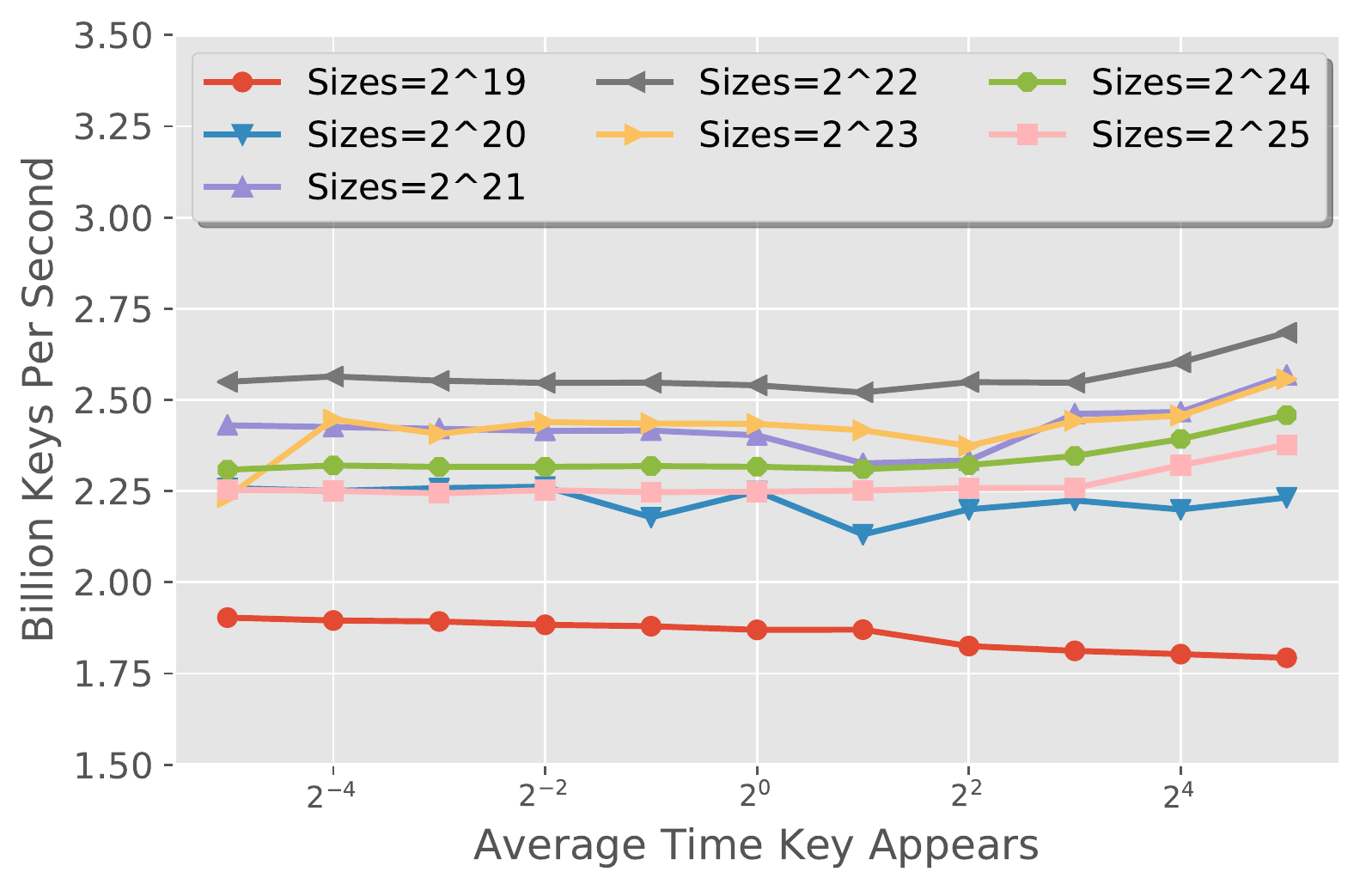}}
\hspace{0.25cm}
\subfloat[Bins - $2^{16}$]{\includegraphics[width=0.60\columnwidth]{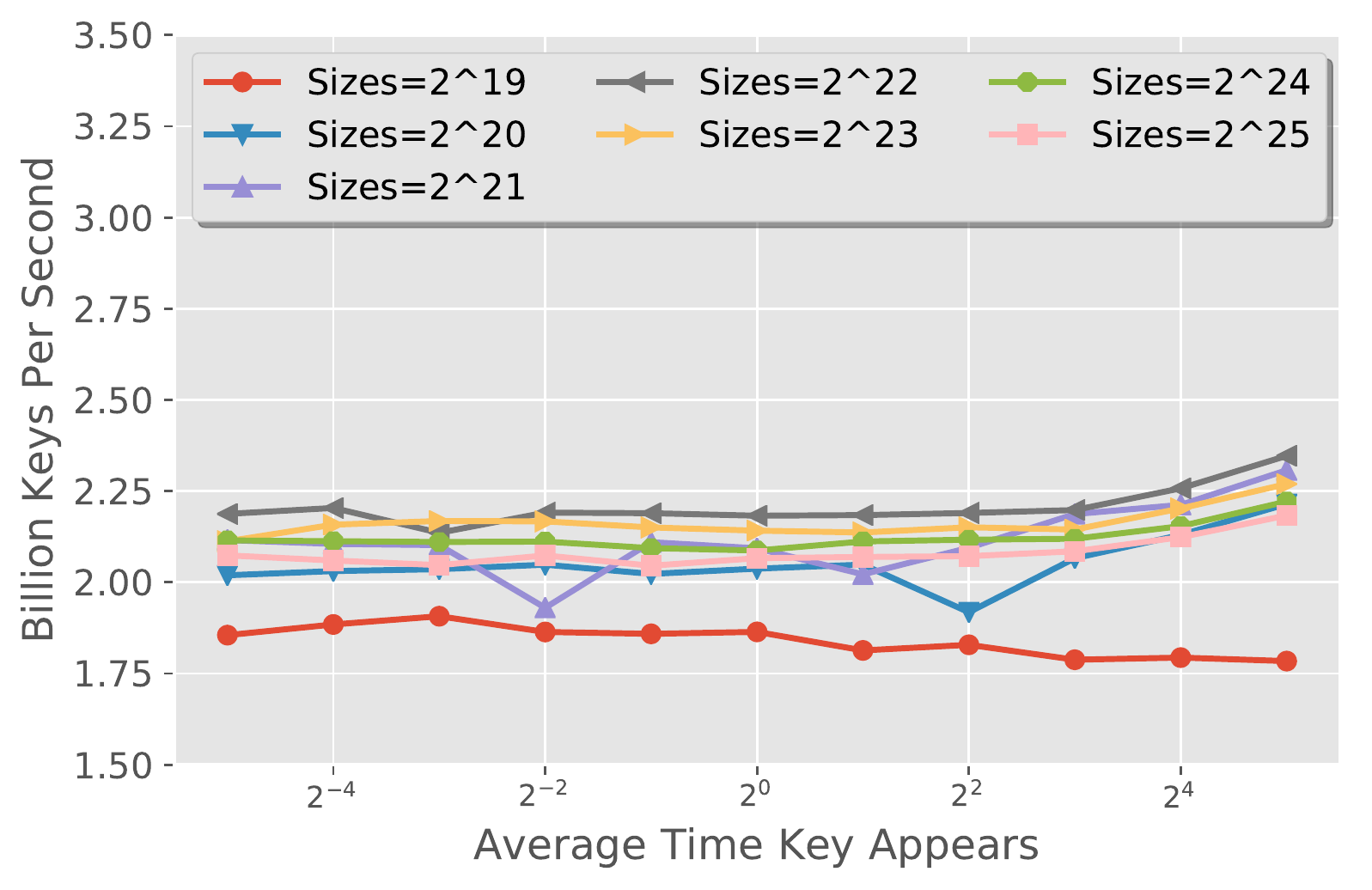}}

\caption{Billions of keys per second of our HashGraph 2.0 as a function of the number of times each key appears in the input. The number of bins is (a) $2^{14}$, (b) $2^{15}$, and (c) $2^{15}$, respectively. The different curves are for different input sizes. On average the best performance of each input size can be found in (a) $2^{14}$ and (b) $2^{15}$.
}

\label{fig:bins-vs-size}

\end{figure*}

\subsection{Hash-Table Building}
\label{sec:perf-hg}
In the following subsection we show the performance of our new HashGraph algorithms for building a hash-table. We show several interesting performance findings with respect to collision and hash-value duplications that are contrary to intuition and the typical performance found in existing hash values. 

\subsubsection*{Framework Comparison}

Fig. \ref{fig:hash-graph} depicts the performance of the various implementations. The input is identical for all the frameworks: $N=2^{25}=32M$ and the input is the sequence $[1,2,3,..,2^{25}]$. 
The x-axis represents the load of the table, such that that the number of entries within the hash table is $\frac{N}{load}$. Larger loads mean denser tables and can impact performance.
{\bf Recall, that the load for HashGraph is different, specifically, the load represents the number of ``vertices'' or unique set of hash values. Also, the number of ``edges'' (the hash-table size) is constant in our approach and is equal to the input size.} 
Fig. \ref{fig:hash-graph} (a) depicts the hash table build in terms of billions of keys per second---higher is better.
Fig. \ref{fig:hash-graph} (b) depicts the speedup of the various algorithms in comparison with cuDPP which is a de-facto hash-table implementation for the GPU.
Key observations:  

\noindent \textbullet \ HashGraph-V2 (Alg. \ref{alg:hg2}) is roughly $4\times$ faster than HashGraph-V1 (Alg. \ref{alg:hg1}).  Recall, that both HashGraph-V1 and HashGraph-V2 both create the same output (though the order of the entries in the hash tables might be different). HashGraph-V2 uses an additional preprocessing phase with radix binning that improves the cache efficiency as there are fewer cache misses.

\noindent \textbullet \ HashGraph-V2 is significantly faster than the other frameworks. Even for the lowest load factor, 0.5,(default  load factor for these frameworks), HashGraph-V2 is faster than cuDPP by as much as $1.6\times$, WarpDrive by $1.8\times$, and cuDF by over $1.3\times$.

\noindent \textbullet \ For higher load factors HashGraph-V2 outperforms cuDPP by as much as $5\times$, WarpDrive by $2.5\times$ times, and cuDF by over $8\times$ times.

\noindent \textbullet \ The performance of HashGraph-V2 actually improves as the load-factor is reduced! This is in contrast with most other hash table building functionality. Note that all of them show reduced performance at load factor $0.9$. For HashGraph-V2 the performance actually increases by roughly $10\%-15\%$. Recall in HashGraph-V2 (Alg. \ref{alg:hg2}) there are several loops that are dependent on the number of vertices (denoted as $C_V$). By increasing the load factor, we in fact decrease the size of $C_V$ (the set of vertices). This leads to fewer iterations for several of the $parallel for$ loops and a faster prefix summation operation. 

\noindent \textbullet \ For HashGraph, we can increase the load factor beyond $1$ to further improve performance (Fig. \ref{fig:probing-load}. Thus, if we compare HashGraph-V2's best performance with the best performance of the other frameworks, HashGraph-V2 is faster than cuDPP by as much as $2\times$, WarpDrive by $2.2\times$ times, and cuDF by over $1.5\times$ times.


\subsubsection*{Collision Management}

A key feature of a hash table is how good it performs when there are collisions in the hash table. In this subsection we evaluate cuDF and HashGraph for various collision rates (which are controlled by the average number of appearances of each input value). We did not collect these numbers for cuDPP and WarpDrive. Specifically, WarpDrive is unable to deal with duplicate keys in the hash table creation phase.

Fig. \ref{fig:dup-keys} depicts the performance, in keys per second, for building a hash table as a function of the average number of times each key appears in the input (going from a small number of times to a large number of times). For cuDF we collected results for two different load factors: $50\%$ and $80\%$. Inputs were created using a uniform random number generator. The x-axis represents the average time each key appears on average in the input. 
For example, given an input with 32M keys and an average of four appearances, the keys are generated from numbers in the range of $[1,..., 8M]$. Note, as the average number of appearances increases, the cost of collision management with open-addressing increased in a quadratic manner. 


In contrast, the performance of HashGraph barely changes with the increase in the number of average values. 
HashGraph offers a new and efficient approach for dealing with collisions with little over head. In comparison with cuDF, when a key appears on average 2, 8, and 32 times, our new HashGraph-V2 approach is $2\times$, $4\times$, and $40\times$  faster, respectively, than the fastest of the two cuDF implementations.

\subsubsection*{Bin Count Selection}

Fig. \ref{fig:size-vs-bins} and Fig. \ref{fig:bins-vs-size} depict the performance of HashGraph-V2 for different bin sizes and input sizes. For both plots, the x-axis depicts the average number of times each elements appears in the input.
Fig. \ref{fig:size-vs-bins} depicts the performance of HashGraph-V2 for different bins sizes. Each of the subplots depicts the performance for a different input sizes: 2M, 8M, and 32M. For all these input sizes, selecting $2^{14}$ or $2^{15}$ bins gives the best performance. Note, that the performance for a given bin count stays constant as the number of average appearances grows. 
Fig. \ref{fig:bins-vs-size} shows the performance of HashGraph-V2 for several different input sizes. 

Intuitively, using a very large number of bins decreases the performance as this becomes very close to HashGraph-V1. In contrast increasing the number of bins beyond a certain point also reduces the performance as it increases hot-spotting and increases serialization of the reads and writes. Another problem of having too many bins is that might not fit in the cache. At the far extreme where the number of bins is equal to the input size, we are once again stuck with random memory accesses (similar to that found in most open-hashing based approaches).


\begin{figure}[t!]

\centering

\includegraphics[width=0.8\columnwidth]{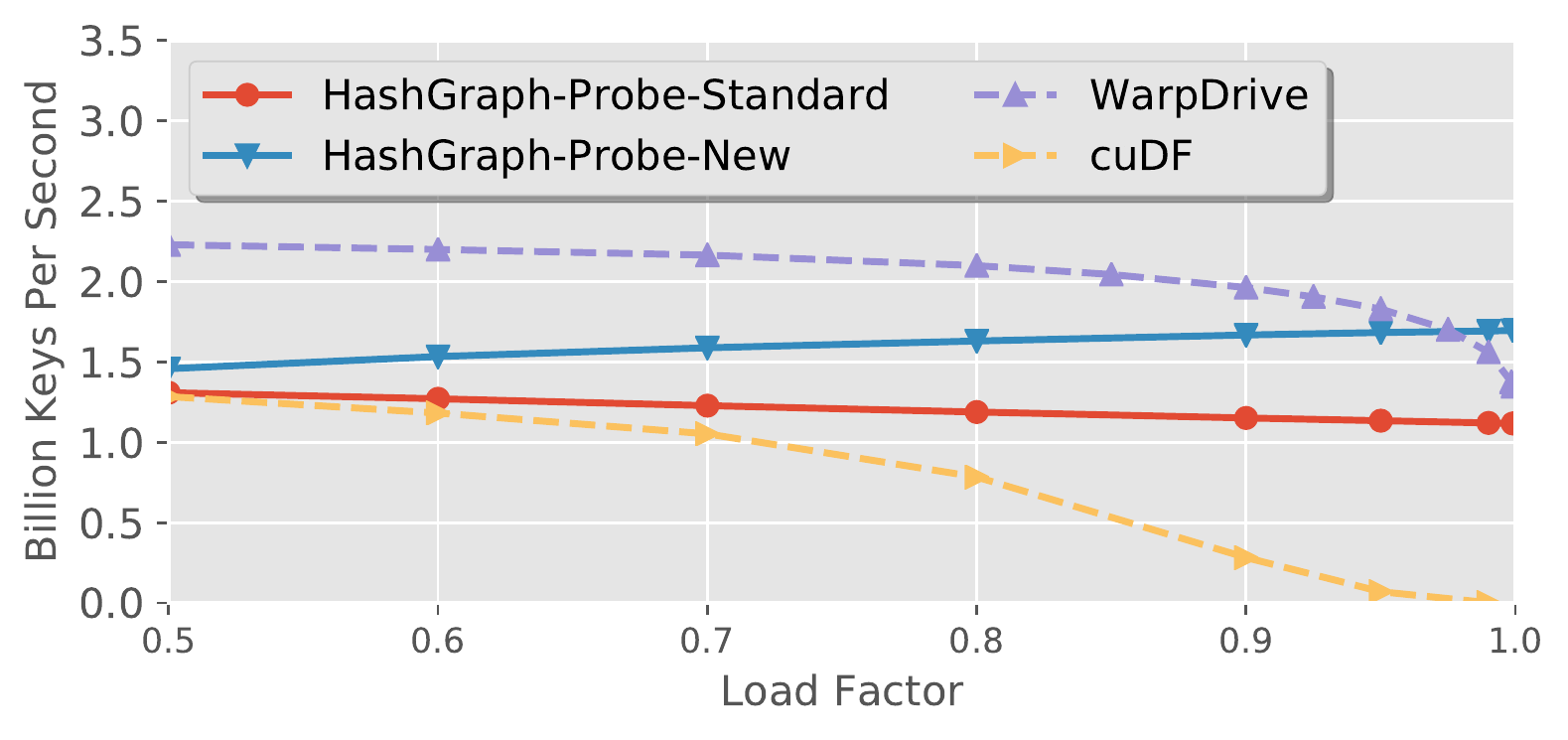}
\caption{Performance of probing. There are two curves for HashGraph, one for a standard probing algorithm that can be also used for open-addressing based tables and the other curve for a new probing that intersects the adjacency lists for the different hash entries. WarpDrive's probing is faster as it finds only the first instance of value whereas the others find all instances.}

\label{fig:probing}

\end{figure}



\begin{figure}[t!]
\centering
\includegraphics[width=0.8\columnwidth]{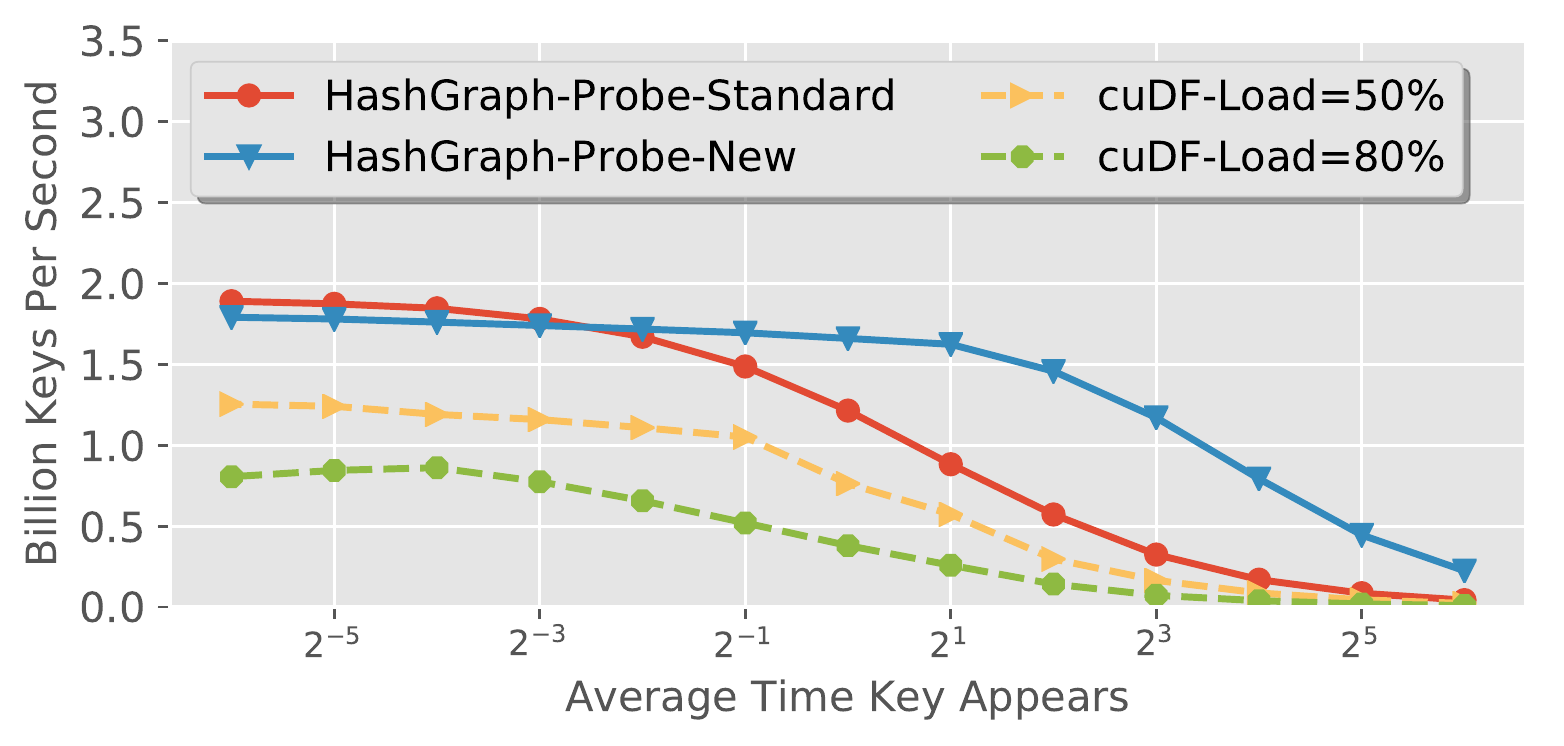}
\caption{Performance comparison of probing of HashGraph with cuDF. The new HashGraph probing algorithm is slightly slower when the number of unique keys is small. However, $10\times$ faster than the standard probing algorithm and roughly $100\times$ faster than cuDF when the number of duplicates is high. }
\label{fig:probing-dist}
\end{figure}

\subsection{Hash-Table Probing}

In this subsection we evaluate the performance of probing. Again to ensure a fair evaluation across the implementations we use the sequence $1,2,..,N$ as our input.
Typically, the probing phase is faster than the table building phase is it only requires random read operations in contrast to random writes (and atomic instructions).



Fig. \ref{fig:probing} depicts the performance of the hash tables as a function of the load factor. For both cuDF and Warpdrive, as the load factor increases hash table becomes denser the performance decreases. 
WarpDrive's probing algorithms finds the first instance of a hashed value and stops; whereas the other implementations find all the instances (and they do a deeper scan). Furthermore, WarpDrive's inability to deal with duplicate key values means that it cannot be used for join operations.
The HashGraph-Probe-Standard probing approach has near constant performance as a function of the load factor. In contrast, for our new probing the approach the performance increases as the load is increased---this is due to the fact that we build a second hash table for the probing input. Recall, this behavior was also seen in the HashGraph building phase. At peak, HashGraph can probe at a rate of 1.8 billion keys per second. 

In an execution breakdown of our the HashGraph-Probe-New probing mechanism, roughly $75\%$ of our probing time is spent building the second hash-table. The remaining time is spent on intersecting the lists. This is highly motivating for two reasons :1) it confirms that our new intersection is fast and cache efficient and 2) if we can further improve the performance of building the hash table we can further improve the performance of the probing.

Fig. \ref{fig:probing-dist} depicts the performance of the probing functionality as a function of the average number of times that each key appears in the input. For both cuDF curves, the performance clearly drops off as the number of appearances increases. This is not surprising as open-addressing becomes more expensive as the table becomes denser and more duplicates are added. In contrast, the probe rate for HashGraph decreases at a much slower rate. Especially for the new approach that has better cache utilization and reuse. The new probing algorithm is roughly $100\times$ faster than cuDF when the number of duplicates is 32.

\begin{figure}[t!]
\centering
\includegraphics[width=0.8\columnwidth]{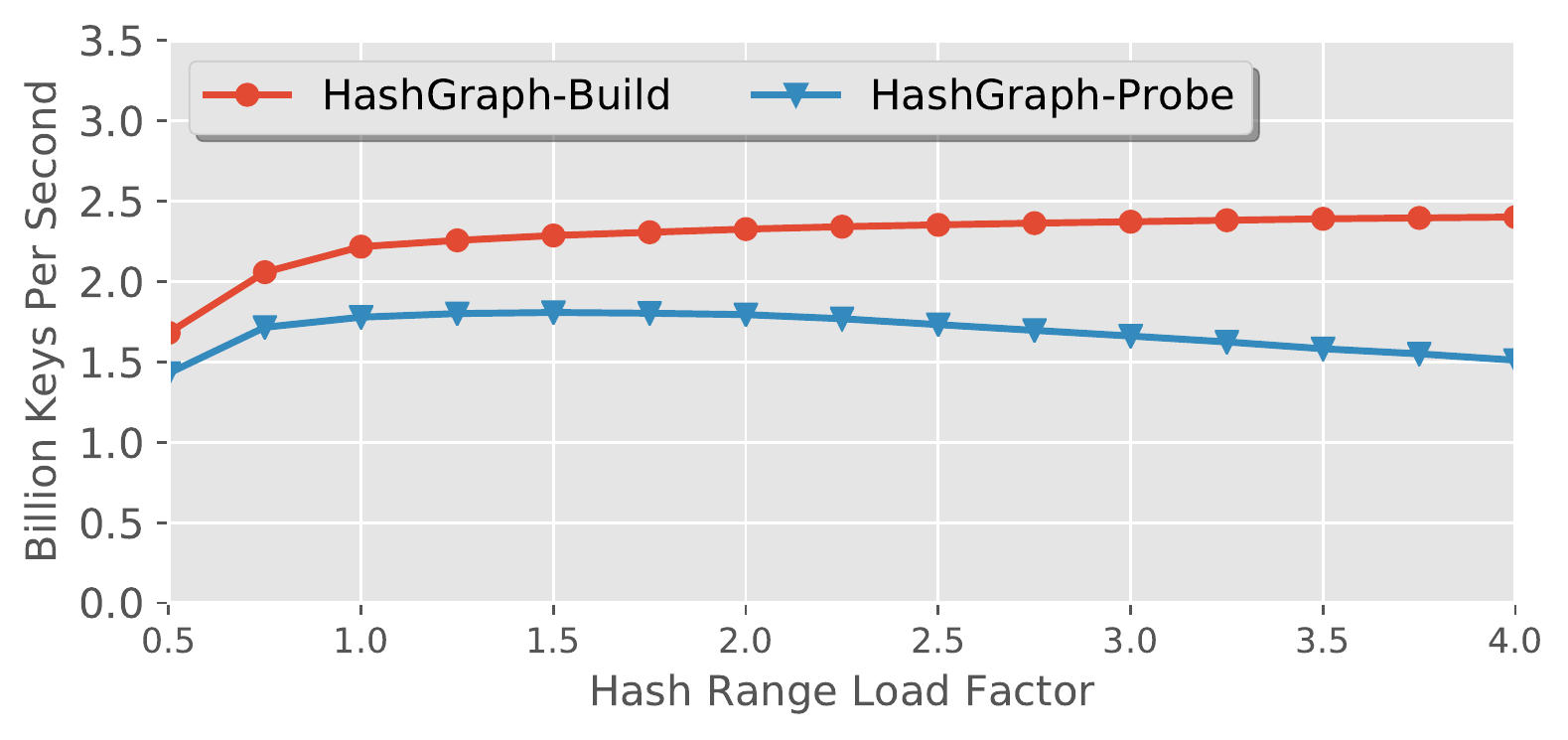}
\caption{Performance comparison of HashGraph as a function of the hash table range, starting on the left where the range is twice as big as the input size and all the way to the right where the range is a quarter of the input size (meaning that there will be many more collisions).
}
\label{fig:probing-load}
\end{figure}

\subsection{Hash Table Ranges, Load Factors, and Vertex Counts}

In this last subsection, we analyze the impact of changing the range of the hash table. Typically this is referred to as the load-factor in open-addressing based approaches and its also well established that decreasing the range increases the number of collisions and reduces the performance. However, for our new HashGraph approach the range refers to the number of vertices (and edge lists) in the HashGraph. And we showed that HashGraph actually gains some performance by reducing the range (vertices) as there are several parallel for loops that are dependent on this size. {\bf Lastly, recall that the number of edges (which are hashed inputs) does not change}. 

Fig. \ref{fig:probing-load} depicts the performance of HashGraph for both the building and probing as a function of the hash range. As the load is increased, from left to right, the effective number of vertices is decreased from being twice the size of the input to being one quarter the size of the input---this is $8\times$ difference in the hash range. The building rate increases from about 1.6 billion keys per second to roughly 2.4 billion keys per second. Though the build rate does not increase much beyond the point that the hash-range is half the size of the input (marked at load factor 2). In contrast, the performance of the probing function peaks within the load-factor of 1 to 2. Beyond this load, the performance of probing goes down as the lists become longer and the hash lists are longer. Within the aforementioned load-factor range there is a good balance between slightly improving hash table building vs. probing. Altogether, it seems that any load between the range of 1 and 2 will give good performance for HashGraph and these can also be decided at runtime based on the properties of the input.

\begin{figure}[t!]
\centering
\includegraphics[width=0.8\columnwidth]{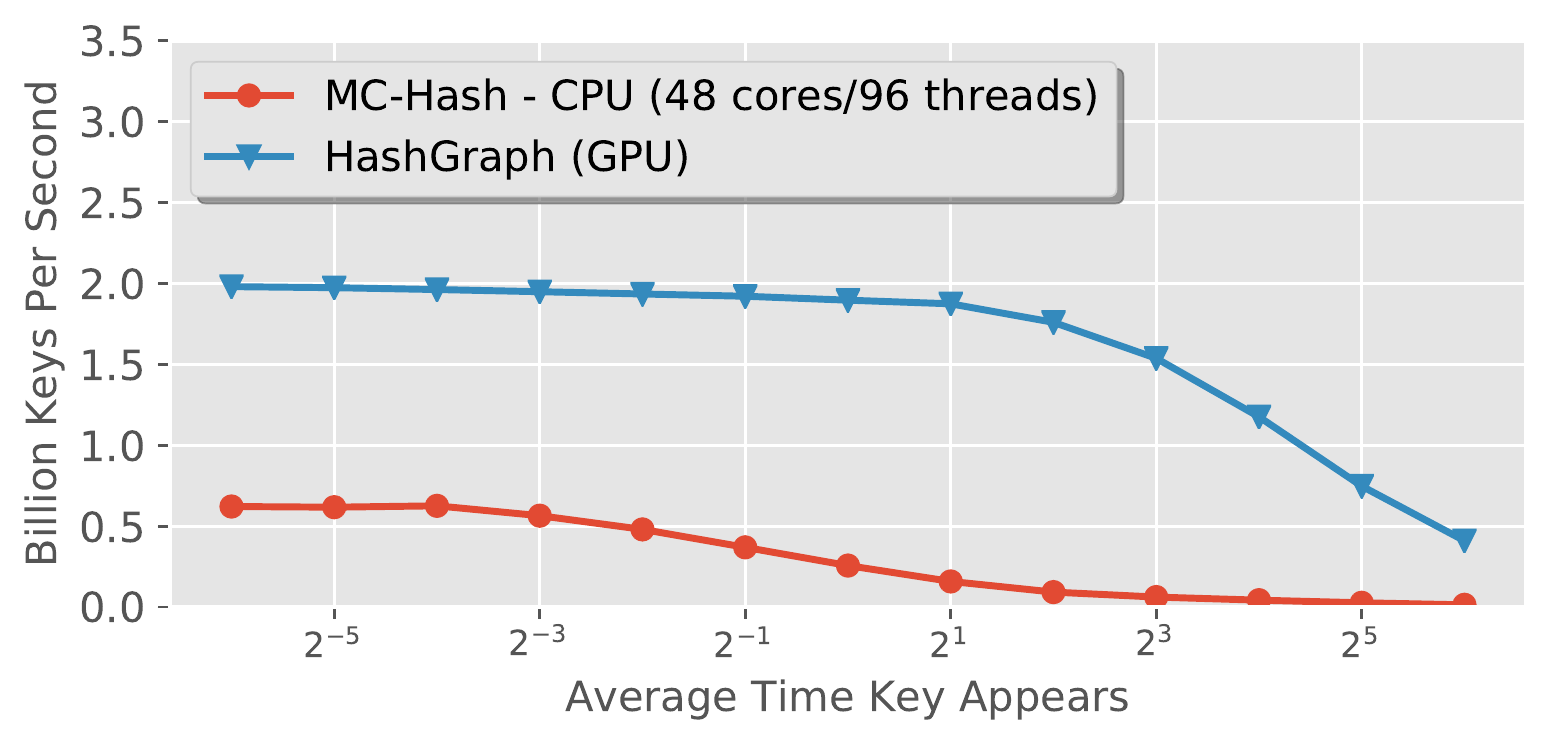}
\caption{Performance comparison (higher is better) of HashGraph with one of the fastest known CPU hash algorithms \cite{balkesen2013main,balkesen2013multi} for an inner-join. The execution time includes both the hash table creation and the probing phases. The lists with the result of the inner join are not actually created or timed. The size of the two inputs is 32M elements.
 }
\label{fig:mc-hash}
\end{figure}

\subsection{HashGraph Vs. CPU HashTables}

In \cite{balkesen2013main,balkesen2013multi} one of the fastest known CPU hash tables implementations is presented. MC-Hash was designed for inner-joins of two relations. MC-Hash times the execution for both building the hash-table as well as its probing with one timer. Using this single timer, we are able to evaluate and compare the performance of this implementation with HashGraph. We use two input arrays of 32M elements that are generated randomly using MC-Hash's generator and benchmark the algorithms with a varying number of average appearance for the keys. To compute the number of keys per second, we take into account the size of both inputs (meaning we diving 64M by the execution time). Performance of MC-Hash and HashGraph for a join operation are depicted in Fig. \ref{fig:mc-hash}. 

For Hash-Graph we calculated the sum of the time for hash-table building and probing. Hash-Graph, using a single GPU, is $3\times$ faster than a 48-core (96-thread) system processor when the number of appearances per key is small. However, when the average number of keys is high, HashGraph outperforms MC-Hash by as much as $10\times$.

Note, the lists with the results of the intersection are not instantiated.The size of such lists would be fairly large especially when the number of duplicates is high. Specifically, for the uniformly randomly generated numbers used in our experiments the output size could easily be one to two orders of magnitude larger than the inputs---especially when the number of duplicates are high. Thus, a lot of time and memory would be spent on creating the lists and would make a comparison of the building and probing more challenging. Further, MC-Hash does not create these list as well.

\section{Summary}

In this paper we presented HashGraph a new approach for both building a hash table and for probing values. The HashGraph approach shows a relationship between building a hash table and creating a sparse graph data structure. 
Unlike past approaches which typically build a hash table in a single sweep over the data, our two new algorithms actually sweep over the data multiple times. By using multiple sweeps, including a pre-processing sweep, we are able to create a very efficient hash table that can deal with collisions in a simple manner. 
Our second algorithm uses an additional sweep over the data that is both cache friendly and bandwidth friendly removing the penalty of random memory accesses. Our second algorithm is anywhere from $50\%-100\%$ faster than several state-of-the-art hash table implementations.

HashGraph shows a new way to manage collisions that takes the best of open-addressing and chaining, without any of their downsides. 
First, we showed that we can place colliding elements in the same spatial locality array without gaps or different hashed values as required in open-addressing or random memory accesses as required with chaining.

Second, using our preprocessing sweep we are able to determine in advance  the exact number of elements that will be placed in each of the so called temporal chains. This removes the need for finding an empty spot in the table. This is important for inputs with a large number of duplicates and overcomes many of the problems with elaborate methods such as Cuckoo hashing.
Thus, we showed a new collision management type for hashing.

Third, the performance of HashGraph is mostly invariant to the key distributions and the number of times each key appears in contrast to most hash table. In the experiment section we showed that the performance of building a table went down by less than $15\%$ even when each key appeared on average $32\times$. Performance for past approaches went down by more than $10\times$.

Fourth, past approaches have typically required roughly $1.5\times-2\times$ more entries than the original input to avoid the problems associated with conflicts. Reducing this storage overhead has typically led to reduced performance. In contrast, HashGraph separates the load factor (which can also be viewed as the number of possible hash entries) with the actual hash table size. With HashGraph the range of the entries can be reduced to the exact number of entries. Surprisingly, decreasing the load factor actually improves the performance of HashGraph. This is in stark contrast to past approaches.


Fifth, we showed a new approach for probing the hash table. Specifically, for list intersections operations (such as those founds in inner joins), we show that we can create a second hash table for both the input lists and then we can intersect the lists of each value in a method that is cache friendly (no random accesses as found in chaining) and also efficient (searches through only relevant values unlike open-addressing).
The uniqueness of this new probing approach is makes this phase more cache friendly. This could be important for future processors where the memory subsystem (bandwidth, cache size, memory channels, latencies) is different.

\subsection*{Dynamic HashGraph}
\label{sec:dynamic-hg}

As mentioned earlier in this paper, hashing is used in a wide range applications. In some applications the hash table is created from a static data set and in other cases the hash table changes by other adding or removing values. These can be referred to as incremental or decremental operations. In this work we primarily focused on creating a table for static data hash tables such as those required for inner join operations and SpGEMM. 

While we do not discuss this in detail, HashGraph can easily support decremental operations simply by storing an additional length array in addition to the offset array. Thus, HashGraph can support deletions whereas open-addressing methods are not able to support deletion without adding a special deletion ``marker'' to the hash table. 

We believe that recent work on developing efficeint dynamic graph data structures might be of great use for HashGraph. While a handful of such data structures have been developed in the last decade \cite{green-hornet,green-custinger,winter2017aims,graphin,macko2015llama}, we believe that Hornet data structure \cite{green-hornet} can be adapted for a dynamic version of HashGraph. This in part due to the fact that Hornet is a dynamic version of CSR.  
We have started to investigate how to implement HashGraph within Hornet and believe that with some modification made to Hornet's initilization process this will possible. Specififically, Hornet initializes process needs to be parallelized as it is currently sequential.
While we have yet to develop a dynamic version of HashGraph, it would seem that the table building process would be the same. They key difference would be is supporting dynamic operations, with an emphasis on insertions.


\subsection*{Conclusions}
In summary, in this paper we presented a novel approach for building and a new approach for probing hash tables. Our analysis includes theoretical complexities as well as empirical performance analysis.
Our new hash table building algorithm is faster than existing state-of-the-art hash tables from $2\times-8\times$ when the number of unique keys is high and up-to $40\times$ faster when the number of unique key is small. Our new probing algorithm can be as much as $100\times$ faster when the number of duplicates is high. We also showed that a single NVIDIA V100 GPU can significantly outperform a 48-core (96 thread) CPU system for an inner-join.

We showed that our the process of building a hash table with HashGraph introduces a new and more efficient collision management that has all the benefits of open-addressing and chaining, but without the drawbacks of these two approaches. We primarily focused on static data hash tables in this paper, though our goal is to investigate how to implement HashGraph for dynamic graphs as well.


\bibliographystyle{ACM-Reference-Format}
\bibliography{bibfile,green,hash}


\begin{thebibliography}{30}


\ifx \showCODEN    \undefined \def \showCODEN     #1{\unskip}     \fi
\ifx \showDOI      \undefined \def \showDOI       #1{#1}\fi
\ifx \showISBNx    \undefined \def \showISBNx     #1{\unskip}     \fi
\ifx \showISBNxiii \undefined \def \showISBNxiii  #1{\unskip}     \fi
\ifx \showISSN     \undefined \def \showISSN      #1{\unskip}     \fi
\ifx \showLCCN     \undefined \def \showLCCN      #1{\unskip}     \fi
\ifx \shownote     \undefined \def \shownote      #1{#1}          \fi
\ifx \showarticletitle \undefined \def \showarticletitle #1{#1}   \fi
\ifx \showURL      \undefined \def \showURL       {\relax}        \fi
\providecommand\bibfield[2]{#2}
\providecommand\bibinfo[2]{#2}
\providecommand\natexlab[1]{#1}
\providecommand\showeprint[2][]{arXiv:#2}

\bibitem[\protect\citeauthoryear{{Albert}, {Jeong}, and
  {Barab{\'a}si}}{{Albert} et~al\mbox{.}}{1999}]%
        {}
\bibfield{author}{\bibinfo{person}{R. {Albert}}, \bibinfo{person}{H. {Jeong}},
  {and} \bibinfo{person}{A.-L. {Barab{\'a}si}}.}
  \bibinfo{year}{1999}\natexlab{}.
\newblock \showarticletitle{{Internet: Diameter of the World-Wide Web}}.
\newblock \bibinfo{journal}{\emph{Nature}}  \bibinfo{volume}{401}
  (\bibinfo{date}{Sept.} \bibinfo{year}{1999}), \bibinfo{pages}{130--131}.
\newblock


\bibitem[\protect\citeauthoryear{Alcantara, Sharf, Abbasinejad, Sengupta,
  Mitzenmacher, Owens, and Amenta}{Alcantara et~al\mbox{.}}{2009}]%
        {alcantara2009real}
\bibfield{author}{\bibinfo{person}{Dan~A Alcantara}, \bibinfo{person}{Andrei
  Sharf}, \bibinfo{person}{Fatemeh Abbasinejad}, \bibinfo{person}{Shubhabrata
  Sengupta}, \bibinfo{person}{Michael Mitzenmacher}, \bibinfo{person}{John~D
  Owens}, {and} \bibinfo{person}{Nina Amenta}.}
  \bibinfo{year}{2009}\natexlab{}.
\newblock \showarticletitle{Real-time parallel hashing on the GPU}.
\newblock \bibinfo{journal}{\emph{ACM Transactions on Graphics (TOG)}}
  \bibinfo{volume}{28}, \bibinfo{number}{5} (\bibinfo{year}{2009}),
  \bibinfo{pages}{154}.
\newblock


\bibitem[\protect\citeauthoryear{Appleby}{Appleby}{2008}]%
        {appleby2008murmurhash}
\bibfield{author}{\bibinfo{person}{Austin Appleby}.}
  \bibinfo{year}{2008}\natexlab{}.
\newblock \bibinfo{title}{Murmurhash 2.0}.
\newblock   (\bibinfo{year}{2008}).
\newblock


\bibitem[\protect\citeauthoryear{Ashkiani, Davidson, Meyer, and Owens}{Ashkiani
  et~al\mbox{.}}{2016}]%
        {ashkiani2016gpu}
\bibfield{author}{\bibinfo{person}{Saman Ashkiani}, \bibinfo{person}{Andrew
  Davidson}, \bibinfo{person}{Ulrich Meyer}, {and} \bibinfo{person}{John~D
  Owens}.} \bibinfo{year}{2016}\natexlab{}.
\newblock \showarticletitle{{GPU Multisplit}}. In \bibinfo{booktitle}{\emph{ACM
  SIGPLAN Notices}}, Vol.~\bibinfo{volume}{51}. ACM, \bibinfo{pages}{12}.
\newblock


\bibitem[\protect\citeauthoryear{Ashkiani, Farach-Colton, and Owens}{Ashkiani
  et~al\mbox{.}}{2018}]%
        {ashkiani2018dynamic}
\bibfield{author}{\bibinfo{person}{Saman Ashkiani}, \bibinfo{person}{Martin
  Farach-Colton}, {and} \bibinfo{person}{John~D Owens}.}
  \bibinfo{year}{2018}\natexlab{}.
\newblock \showarticletitle{A dynamic hash table for the GPU}. In
  \bibinfo{booktitle}{\emph{2018 IEEE International Parallel and Distributed
  Processing Symposium (IPDPS)}}. IEEE, \bibinfo{pages}{419--429}.
\newblock


\bibitem[\protect\citeauthoryear{Balkesen, Alonso, Teubner, and
  {\"O}zsu}{Balkesen et~al\mbox{.}}{2013a}]%
        {balkesen2013multi}
\bibfield{author}{\bibinfo{person}{Cagri Balkesen}, \bibinfo{person}{Gustavo
  Alonso}, \bibinfo{person}{Jens Teubner}, {and} \bibinfo{person}{M~Tamer
  {\"O}zsu}.} \bibinfo{year}{2013}\natexlab{a}.
\newblock \showarticletitle{{Multi-Core, Main-Memory Joins: Sort vs. Hash
  Revisited}}.
\newblock \bibinfo{journal}{\emph{Proceedings of the VLDB Endowment}}
  \bibinfo{volume}{7}, \bibinfo{number}{1} (\bibinfo{year}{2013}),
  \bibinfo{pages}{85--96}.
\newblock


\bibitem[\protect\citeauthoryear{Balkesen, Teubner, Alonso, and
  {\"O}zsu}{Balkesen et~al\mbox{.}}{2013b}]%
        {balkesen2013main}
\bibfield{author}{\bibinfo{person}{Cagri Balkesen}, \bibinfo{person}{Jens
  Teubner}, \bibinfo{person}{Gustavo Alonso}, {and} \bibinfo{person}{M~Tamer
  {\"O}zsu}.} \bibinfo{year}{2013}\natexlab{b}.
\newblock \showarticletitle{{Main-Memory Hash Joins on Multi-Core CPUs: Tuning
  to the Underlying Hardware}}. In \bibinfo{booktitle}{\emph{IEEE 29th
  International Conference on Data Engineering (ICDE)}}. IEEE,
  \bibinfo{pages}{362--373}.
\newblock


\bibitem[\protect\citeauthoryear{Barthels, M{\"u}ller, Schneider, Alonso, and
  Hoefler}{Barthels et~al\mbox{.}}{2017}]%
        {barthels2017distributed}
\bibfield{author}{\bibinfo{person}{Claude Barthels}, \bibinfo{person}{Ingo
  M{\"u}ller}, \bibinfo{person}{Timo Schneider}, \bibinfo{person}{Gustavo
  Alonso}, {and} \bibinfo{person}{Torsten Hoefler}.}
  \bibinfo{year}{2017}\natexlab{}.
\newblock \showarticletitle{{Distributed Join Algorithms on Thousands of
  Cores}}.
\newblock \bibinfo{journal}{\emph{Proceedings of the VLDB Endowment}}
  \bibinfo{volume}{10}, \bibinfo{number}{5} (\bibinfo{year}{2017}),
  \bibinfo{pages}{517--528}.
\newblock


\bibitem[\protect\citeauthoryear{Bethel, Gosink, Wu, Bethel, Owens, and
  Joy}{Bethel et~al\mbox{.}}{2008}]%
        {bethel2008bin}
\bibfield{author}{\bibinfo{person}{Edward~W Bethel}, \bibinfo{person}{Luke~J
  Gosink}, \bibinfo{person}{Kesheng Wu}, \bibinfo{person}{Edward~Wes Bethel},
  \bibinfo{person}{John~D Owens}, {and} \bibinfo{person}{Kenneth~I Joy}.}
  \bibinfo{year}{2008}\natexlab{}.
\newblock \bibinfo{booktitle}{\emph{{Bin-Hash Indexing: A Parallel Method for
  Fast Query Processing}}}.
\newblock \bibinfo{type}{{T}echnical {R}eport}. \bibinfo{institution}{Lawrence
  Berkeley National Lab.(LBNL), Berkeley, CA (United States)}.
\newblock


\bibitem[\protect\citeauthoryear{Bisson and Fatica}{Bisson and Fatica}{2017}]%
        {bisson2017high}
\bibfield{author}{\bibinfo{person}{Mauro Bisson} {and}
  \bibinfo{person}{Massimiliano Fatica}.} \bibinfo{year}{2017}\natexlab{}.
\newblock \showarticletitle{High Performance Exact Triangle Counting on GPUs}.
\newblock \bibinfo{journal}{\emph{IEEE Transactions on Parallel and Distributed
  Systems}} \bibinfo{volume}{28}, \bibinfo{number}{12} (\bibinfo{year}{2017}),
  \bibinfo{pages}{3501--3510}.
\newblock


\bibitem[\protect\citeauthoryear{Blanas, Li, and Patel}{Blanas
  et~al\mbox{.}}{2011}]%
        {blanas2011design}
\bibfield{author}{\bibinfo{person}{Spyros Blanas}, \bibinfo{person}{Yinan Li},
  {and} \bibinfo{person}{Jignesh~M Patel}.} \bibinfo{year}{2011}\natexlab{}.
\newblock \showarticletitle{{Design and Evaluation of Main Memory Hash Join
  Algorithms for Multi-Core CPUs}}. In \bibinfo{booktitle}{\emph{Proceedings of
  the 2011 ACM SIGMOD International Conference on Management of data}}. ACM,
  \bibinfo{pages}{37--48}.
\newblock


\bibitem[\protect\citeauthoryear{Blelloch}{Blelloch}{1990}]%
        {blelloch1990pre}
\bibfield{author}{\bibinfo{person}{Guy~E Blelloch}.}
  \bibinfo{year}{1990}\natexlab{}.
\newblock \bibinfo{booktitle}{\emph{{Prefix Sums and Their Applications}}}.
\newblock \bibinfo{type}{{T}echnical {R}eport}.
  \bibinfo{institution}{Citeseer}.
\newblock


\bibitem[\protect\citeauthoryear{Busato, Green, Bombieri, and Bader}{Busato
  et~al\mbox{.}}{2018}]%
        {green-hornet}
\bibfield{author}{\bibinfo{person}{F. Busato}, \bibinfo{person}{O. Green},
  \bibinfo{person}{N. Bombieri}, {and} \bibinfo{person}{D.A. Bader}.}
  \bibinfo{year}{2018}\natexlab{}.
\newblock \showarticletitle{{Hornet: An Efficient Data Structure for Dynamic
  Sparse Graphs and Matrices on GPUs}}. In \bibinfo{booktitle}{\emph{IEEE
  Proc.\ High Performance Extreme Computing ({HPEC})}}.
  \bibinfo{address}{Waltham, MA}.
\newblock


\bibitem[\protect\citeauthoryear{Green and Bader}{Green and Bader}{2016}]%
        {green-custinger}
\bibfield{author}{\bibinfo{person}{O. Green} {and} \bibinfo{person}{D.A.
  Bader}.} \bibinfo{year}{2016}\natexlab{}.
\newblock \showarticletitle{{cuSTINGER: Supporting Dynamic Graph Algorithms for
  GPUS}}. In \bibinfo{booktitle}{\emph{IEEE Proc.\ High Performance Extreme
  Computing ({HPEC})}}. \bibinfo{address}{Waltham, MA}.
\newblock


\bibitem[\protect\citeauthoryear{Harris, Owens, Sengupta, Zhang, and
  Davidson}{Harris et~al\mbox{.}}{2007a}]%
        {harris2007cudpp}
\bibfield{author}{\bibinfo{person}{Mark Harris}, \bibinfo{person}{John Owens},
  \bibinfo{person}{Shubho Sengupta}, \bibinfo{person}{Yao Zhang}, {and}
  \bibinfo{person}{Andrew Davidson}.} \bibinfo{year}{2007}\natexlab{a}.
\newblock \bibinfo{title}{{CUDPP: CUDA Data Parallel Primitives library}}.
\newblock   (\bibinfo{year}{2007}).
\newblock


\bibitem[\protect\citeauthoryear{Harris, Sengupta, and Owens}{Harris
  et~al\mbox{.}}{2007b}]%
        {harris2007parallel}
\bibfield{author}{\bibinfo{person}{Mark Harris}, \bibinfo{person}{Shubhabrata
  Sengupta}, {and} \bibinfo{person}{John~D Owens}.}
  \bibinfo{year}{2007}\natexlab{b}.
\newblock \showarticletitle{{Parallel prefix sum (scan) with CUDA}}.
\newblock \bibinfo{journal}{\emph{{GPU} gems}} \bibinfo{volume}{3},
  \bibinfo{number}{39} (\bibinfo{year}{2007}), \bibinfo{pages}{851--876}.
\newblock


\bibitem[\protect\citeauthoryear{J{\"u}nger, Hundt, and Schmidt}{J{\"u}nger
  et~al\mbox{.}}{2018}]%
        {junger2018warpdrive}
\bibfield{author}{\bibinfo{person}{Daniel J{\"u}nger},
  \bibinfo{person}{Christian Hundt}, {and} \bibinfo{person}{Bertil Schmidt}.}
  \bibinfo{year}{2018}\natexlab{}.
\newblock \showarticletitle{{WarpDrive: Massively Parallel Hashing on Multi-GPU
  Nodes}}. In \bibinfo{booktitle}{\emph{2018 IEEE International Parallel and
  Distributed Processing Symposium (IPDPS)}}. IEEE, \bibinfo{pages}{441--450}.
\newblock


\bibitem[\protect\citeauthoryear{Karnagel, Mueller, and Lohman}{Karnagel
  et~al\mbox{.}}{2015}]%
        {karnagel2015optimizing}
\bibfield{author}{\bibinfo{person}{Tomas Karnagel}, \bibinfo{person}{Rene
  Mueller}, {and} \bibinfo{person}{Guy~M Lohman}.}
  \bibinfo{year}{2015}\natexlab{}.
\newblock \showarticletitle{{Optimizing GPU-accelerated Group-By and
  Aggregation.}}
\newblock \bibinfo{journal}{\emph{ADMS at VLDB}}  \bibinfo{volume}{8}
  (\bibinfo{year}{2015}), \bibinfo{pages}{20}.
\newblock


\bibitem[\protect\citeauthoryear{Khorasani, Belviranli, Gupta, and
  Bhuyan}{Khorasani et~al\mbox{.}}{2015}]%
        {khorasani2015stadium}
\bibfield{author}{\bibinfo{person}{Farzad Khorasani}, \bibinfo{person}{Mehmet~E
  Belviranli}, \bibinfo{person}{Rajiv Gupta}, {and} \bibinfo{person}{Laxmi~N
  Bhuyan}.} \bibinfo{year}{2015}\natexlab{}.
\newblock \showarticletitle{{Stadium Hashing: Scalable and Flexible Hashing on
  Gpus}}. In \bibinfo{booktitle}{\emph{International Conference on Parallel
  Architecture and Compilation (PACT)}}. IEEE, \bibinfo{pages}{63--74}.
\newblock


\bibitem[\protect\citeauthoryear{Kim, Kaldewey, Lee, Sedlar, Nguyen, Satish,
  Chhugani, Di~Blas, and Dubey}{Kim et~al\mbox{.}}{2009}]%
        {kim2009sort}
\bibfield{author}{\bibinfo{person}{Changkyu Kim}, \bibinfo{person}{Tim
  Kaldewey}, \bibinfo{person}{Victor~W Lee}, \bibinfo{person}{Eric Sedlar},
  \bibinfo{person}{Anthony~D Nguyen}, \bibinfo{person}{Nadathur Satish},
  \bibinfo{person}{Jatin Chhugani}, \bibinfo{person}{Andrea Di~Blas}, {and}
  \bibinfo{person}{Pradeep Dubey}.} \bibinfo{year}{2009}\natexlab{}.
\newblock \showarticletitle{{Sort vs. Hash Revisited: Fast Join Implementation
  on Modern Multi-Core CPUs}}.
\newblock \bibinfo{journal}{\emph{Proceedings of the VLDB Endowment}}
  \bibinfo{volume}{2}, \bibinfo{number}{2} (\bibinfo{year}{2009}),
  \bibinfo{pages}{1378--1389}.
\newblock


\bibitem[\protect\citeauthoryear{Li, Shrivastava, and Konig}{Li
  et~al\mbox{.}}{2012}]%
        {li2012gpu}
\bibfield{author}{\bibinfo{person}{Ping Li}, \bibinfo{person}{Anshumali
  Shrivastava}, {and} \bibinfo{person}{Christian~A Konig}.}
  \bibinfo{year}{2012}\natexlab{}.
\newblock \showarticletitle{{GPU-based Minwise Hashing}}. In
  \bibinfo{booktitle}{\emph{Proceedings of the 21st International Conference on
  World Wide Web}}. ACM, \bibinfo{pages}{565--566}.
\newblock


\bibitem[\protect\citeauthoryear{Macko, Marathe, Margo, and Seltzer}{Macko
  et~al\mbox{.}}{2015}]%
        {macko2015llama}
\bibfield{author}{\bibinfo{person}{Peter Macko}, \bibinfo{person}{Virendra~J
  Marathe}, \bibinfo{person}{Daniel~W Margo}, {and} \bibinfo{person}{Margo~I
  Seltzer}.} \bibinfo{year}{2015}\natexlab{}.
\newblock \showarticletitle{{{LLAMA}: Efficient Graph Analytics Using Large
  Multiversioned Arrays}}. In \bibinfo{booktitle}{\emph{31st IEEE Int'l Conf.
  on Data Engineering (ICDE)}}. \bibinfo{pages}{363--374}.
\newblock


\bibitem[\protect\citeauthoryear{Maier, Sanders, and Dementiev}{Maier
  et~al\mbox{.}}{2016}]%
        {maier2016concurrent}
\bibfield{author}{\bibinfo{person}{Tobias Maier}, \bibinfo{person}{Peter
  Sanders}, {and} \bibinfo{person}{Roman Dementiev}.}
  \bibinfo{year}{2016}\natexlab{}.
\newblock \showarticletitle{{Concurrent Hash Tables: Fast and General?(!)}}. In
  \bibinfo{booktitle}{\emph{ACM SIGPLAN Notices}}, Vol.~\bibinfo{volume}{51}.
  ACM, \bibinfo{pages}{34}.
\newblock


\bibitem[\protect\citeauthoryear{{NVIDIA}}{{NVIDIA}}{2018}]%
        {cudf}
\bibfield{author}{\bibinfo{person}{{NVIDIA}}.} \bibinfo{year}{2018}\natexlab{}.
\newblock \bibinfo{title}{CUDF}.
\newblock   (\bibinfo{year}{2018}).
\newblock


\bibitem[\protect\citeauthoryear{Pagh and Rodler}{Pagh and Rodler}{2004}]%
        {pagh2004cuckoo}
\bibfield{author}{\bibinfo{person}{Rasmus Pagh} {and}
  \bibinfo{person}{Flemming~Friche Rodler}.} \bibinfo{year}{2004}\natexlab{}.
\newblock \showarticletitle{{Cuckoo Hashing}}.
\newblock \bibinfo{journal}{\emph{Journal of Algorithms}} \bibinfo{volume}{51},
  \bibinfo{number}{2} (\bibinfo{year}{2004}), \bibinfo{pages}{122--144}.
\newblock


\bibitem[\protect\citeauthoryear{Pan, Misra, and Aluru}{Pan
  et~al\mbox{.}}{2018}]%
        {pan2018optimizing}
\bibfield{author}{\bibinfo{person}{Tony~C Pan}, \bibinfo{person}{Sanchit
  Misra}, {and} \bibinfo{person}{Srinivas Aluru}.}
  \bibinfo{year}{2018}\natexlab{}.
\newblock \showarticletitle{{Optimizing High Performance Distributed Memory
  Parallel Hash Tables for DNA k-mer Counting}}. In
  \bibinfo{booktitle}{\emph{Optimizing High Performance Distributed Memory
  Parallel Hash Tables for DNA k-mer Counting}}. IEEE, \bibinfo{pages}{0}.
\newblock


\bibitem[\protect\citeauthoryear{Richa, Mitzenmacher, and Sitaraman}{Richa
  et~al\mbox{.}}{2001}]%
        {richa2001power}
\bibfield{author}{\bibinfo{person}{Andrea~W Richa}, \bibinfo{person}{M
  Mitzenmacher}, {and} \bibinfo{person}{R Sitaraman}.}
  \bibinfo{year}{2001}\natexlab{}.
\newblock \showarticletitle{{The Power of Two Random Choices: A Survey of
  Techniques and Results}}.
\newblock \bibinfo{journal}{\emph{Combinatorial Optimization}}
  \bibinfo{volume}{9} (\bibinfo{year}{2001}), \bibinfo{pages}{255--304}.
\newblock


\bibitem[\protect\citeauthoryear{Sengupta, Sundaram, Zhu, Willke, Young, Wolf,
  and Schwan}{Sengupta et~al\mbox{.}}{2016}]%
        {graphin}
\bibfield{author}{\bibinfo{person}{Dipanjan Sengupta},
  \bibinfo{person}{Narayanan Sundaram}, \bibinfo{person}{Xia Zhu},
  \bibinfo{person}{Theodore~L Willke}, \bibinfo{person}{Jeffrey Young},
  \bibinfo{person}{Matthew Wolf}, {and} \bibinfo{person}{Karsten Schwan}.}
  \bibinfo{year}{2016}\natexlab{}.
\newblock \showarticletitle{{Graphin: An Online High Performance Incremental
  Graph Processing Framework}}. In \bibinfo{booktitle}{\emph{European
  Conference on Parallel Processing}}. Springer, \bibinfo{pages}{319--333}.
\newblock


\bibitem[\protect\citeauthoryear{Sengupta, Harris, Zhang, and Owens}{Sengupta
  et~al\mbox{.}}{2007}]%
        {sengupta2007scan}
\bibfield{author}{\bibinfo{person}{Shubhabrata Sengupta}, \bibinfo{person}{Mark
  Harris}, \bibinfo{person}{Yao Zhang}, {and} \bibinfo{person}{John~D Owens}.}
  \bibinfo{year}{2007}\natexlab{}.
\newblock \showarticletitle{{Scan Primitives for GPU Computing}}. In
  \bibinfo{booktitle}{\emph{Graphics hardware}}. \bibinfo{pages}{97--106}.
\newblock


\bibitem[\protect\citeauthoryear{Winter, Zayer, and Steinberger}{Winter
  et~al\mbox{.}}{2017}]%
        {winter2017aims}
\bibfield{author}{\bibinfo{person}{Martin Winter}, \bibinfo{person}{Rhaleb
  Zayer}, {and} \bibinfo{person}{Markus Steinberger}.}
  \bibinfo{year}{2017}\natexlab{}.
\newblock \showarticletitle{{Autonomous, Independent Management of Dynamic
  Graphs on GPUs}}. In \bibinfo{booktitle}{\emph{International Supercomputing
  Conference}}. Springer, \bibinfo{pages}{97--119}.
\newblock


\end{thebibliography}




\end{document}